\documentclass[useAMS,usenatbib]{mn2e}
\usepackage{graphicx,times,array,ctable,amsmath,amssymb}

\newcommand{\Msun}{\mathrm{M}_{\odot}}

\newcommand{\da}{\delta}
\newcommand{\dd}{\mathrm{d}}
\newcommand{\pd}{\partial}
\newcommand{\coco}{ \langle \da_{i} \da_{j} \da_{k}\rangle_c }

\newcommand{\Wgm}{W^{\mathrm{gm}}}
\newcommand{\PIgm}{\Pi_{\epsilon}^{\mathrm{gm}}}
\newcommand{\PIgmF}{\Pi_{\epsilon=0}^{\mathrm{gm}}}

\newcommand{\erf}{\mathrm{erf}}                                                                                                       

\newcommand{\fNL}{f_{\mathrm{NL}}}
\newcommand{\MM}{\mathcal{M}}
\newcommand{\fcoll}{f_{\mathrm{coll}}}
\newcommand{\avgfcoll}{f_{\mathrm{coll,0}}}
\newcommand{\fcollgm}{f_{\mathrm{coll}}^{\mathrm{gm}}}
\newcommand{\avgfcollgm}{f_{\mathrm{coll,0}}^{\mathrm{gm}}}

\def\intc{{\int_{-\infty}^{\delta_c}}}
\def\ddl{{d\delta_1\cdots d\delta_{l-1}}}

\def\ddss{{d\delta_{l+1}\cdots d\delta_{s-1}}}

\def\Wls{{W_{l,s}^{gm}}}
\def\Wl{{W_{0,l}^{gm}}}

\def\dc{{\delta_c}}
\def\dl{{\delta_l}}
\def\dcl{{\dc-\dl}}

\def\ss{{S_s}}
\def\sl{{S_l}}
\def\ssl{{\ss-\sl}}

\def\Ue{{U_{\epsilon}(l,s)}}
\def\Pe{{\Pi_{\epsilon}^{gm}(0,l)}}

\newcommand{\Afd}{\mathcal{A}}
\newcommand{\Bfd}{\mathcal{B}}
\newcommand{\Cfd}{\mathcal{C}}

\title[Halo statistics in non-Gaussian cosmologies]{Halo statistics in non-Gaussian cosmologies:  the collapsed fraction, conditional mass function, and halo bias from the path-integral excursion set method.}

\author[D'Aloisio, Zhang, Jeong, and Shapiro]{Anson
  D'Aloisio$^{1}$\thanks{Email: anson@astro.as.utexas.edu}, 
	Jun Zhang$^{1}$, Donghui Jeong$^{2}$ and Paul R. Shapiro$^{1}$ \\
$^1$Department of Astronomy and Texas Cosmology Center, University of Texas, Austin, TX 78712, USA\\
$^2$Department of Physics and Astronomy, Johns Hopkins University, 3400 N. Charles St., Baltimore, MD 21210}

\begin{document}

\onecolumn

\maketitle

\begin{abstract}
Characterizing the level of primordial non-Gaussianity (PNG) in the initial conditions for structure formation is one of the most promising ways to test inflation and differentiate among different scenarios.  The scale-dependent imprint of PNG on the large-scale clustering of galaxies and quasars has already been used to place significant constraints on the level of PNG in our observed Universe.  Such measurements depend upon an accurate and robust theory of how PNG affects the bias of galactic halos relative to the underlying matter density field.  We improve upon previous work by employing a more general analytical method  - the path-integral extension of the excursion set formalism - which is able to account for the non-Markovianity caused by PNG in the random-walk model used to identify halos in the initial density field.   This non-Markovianity encodes information about environmental effects on halo formation which have so far not been taken into account in analytical bias calculations.  We compute both scale-dependent and -independent corrections to the halo bias, along the way presenting an expression for the conditional collapsed fraction for the first time, and a new expression for the conditional halo mass function.  To leading order in our perturbative calculation, we recover the halo bias results of Desjacques et. al. (2011), including the new scale-dependent correction reported there.  However, we show that the non-Markovian dynamics from PNG can lead to marked differences in halo bias when next-to-leading order terms are included.  We quantify these differences here.  We find that the next-to-leading order corrections suppress the amplitudes of both the scale-dependent and -independent bias by $\sim5-10\%$ for massive halos with $M\sim 10^{15} \Msun/h$, and $\sim30-40\%$ for halos with $M\sim 10^{14} \Msun/h$.  The corrections appear to be more significant as the halo mass is lowered, though we caution that the apparently large effects we observe in the low-mass regime likely signal a breakdown of the perturbative approach taken here.          
\end{abstract}


\section{Introduction}
The inflationary paradigm provides a robust framework for explaining key aspects of our observable Universe such as its geometric flatness, features of the cosmic microwave background, and the initial conditions for structure formation.  Despite these remarkable successes, we still know very little about the physics behind it, and currently cannot distinguish among a wide variety of viable inflationary models.  One of the most promising ways to differentiate among these models is  to probe the statistics of the initial density fluctuations \citep[see][and references therein]{2010CQGra..27l4011D}.  While the simplest scenarios - the canonical single-field slow-roll models - predict an almost perfectly Gaussian distribution of initial fluctuations,  more general inflationary models predict significant deviations from Gaussianity that observations might yet be sensitive enough to detect.  There is therefore great interest in developing ways to measure primordial non-Gaussianity (PNG) since its detection would have profound implications for inflationary theory. 

There are presently two methods that have so far been applied with some success to place significant constraints on the level of PNG in our observed Universe.  The first is the statistical imprint of PNG in the temperature anisotropies of the cosmic microwave background (CMB), which directly probe the initial fluctuations while they are still in the linear regime \citep[e.g.][]{2010CQGra..27l4010K}.  The second is the imprint on the large-scale structure that develops as the initial fluctuations grow to the highly non-linear point of forming galaxies, which serve as tracers of the underlying matter.  In general, non-linear growth can complicate the interpretation of the observed structure, both because density fluctuations develop non-Gaussianity even when the initial conditions are purely Gaussian, and because the theoretical predictions in this regime ultimately depend upon N-body simulations.  On large enough scales, however, the density fluctuations filtered on these scales are still linear, and the problem reduces to the theory of how well galaxies trace the large-scale mass distribution - the so-called ``bias."  

The prospect of probing PNG with large-scale structure improved dramatically when it was discovered that the mode coupling effects of PNG induce a scale-dependent signature in the power spectrum of biased tracers (such as galactic halos) on large scales \citep{2008PhRvD..77l3514D, 2008ApJ...677L..77M,2008PhRvD..78l3507A}.  The first and best studied example of this signature involves the local-quadratic model of PNG, in which the primordial Bardeen potential fluctuation in the matter-dominated epoch, $\Phi_{\mathrm{NG}}(\boldsymbol{x})$, is obtained from a quadratic transformation of the local Gaussian fluctuation field, $\phi_{\mathrm{G}}(\boldsymbol{x})$, according to

\begin{equation}
\Phi_{\mathrm{NG}}(\boldsymbol{x}) = \phi_{\mathrm{G}}(\boldsymbol{x}) + \fNL \left[  \phi_{\mathrm{G}}(\boldsymbol{x})^2 - \langle \phi_{\mathrm{G}}^2(\boldsymbol{x}) \rangle \right],
\label{EQ:localquadratic}
\end{equation}
where $\fNL$ is the so-called non-linearity parameter \citep{1990PhRvD..42.3936S,2001PhRvD..63f3002K}.  In this case, while the Gaussian field is entirely characterized by its power spectrum, $P_{\Phi}(k)$, information about higher-order correlations, for example through the bispectrum, is required to characterize the statistics of the non-Gaussian potential.  To first order in $\fNL$, equation (\ref{EQ:localquadratic}) gives a bispectrum with the form\footnote{The bispectrum in equation (\ref{EQ:localquadraticbispectrum}) is more general than equation (\ref{EQ:localquadratic}), as it can be generated in number of different models that do not involve the latter \citep[see footnote 34 in][for example]{2011ApJS..192...18K}. It is nonetheless customary to refer to this form for the bispectrum as the local template. },

\begin{align}
B^{\mathrm{local}}_{\Phi}(k_1,k_2,k_3) = 2 \fNL [ P_{\Phi}(k_1)P_{\Phi}(k_2) + P_{\Phi}(k_1)P_{\Phi}(k_3) + P_{\Phi}(k_2)P_{\Phi}(k_3) ].
\label{EQ:localquadraticbispectrum}
\end{align}
 Observational constraints have been placed on this form of the bispectrum by finding the range of allowed amplitudes, expressed in terms of $\fNL$.  For example, the CMB anisotropy measurements by the Wilkinson Microwave Anisotropy Probe seven-year data analysis (WMAP7) find a $95\%$ limit of $-10 < \fNL < 74$ \citep{2011ApJS..192...18K}.  The bispectrum in equation (\ref{EQ:localquadraticbispectrum}) is important in the phenomenology of PNG because a detection of non-zero $\fNL$ would rule out standard single-field inflation \citep{2004JCAP...10..006C,2005JCAP...06..003S,2007JCAP...01..002C,2008JCAP...02..021C}.

In order to constrain equation (\ref{EQ:localquadraticbispectrum}) from observations of large-scale structure, it is necessary to consider the expected halo bias for this model, i.e. the ratio of the fractional halo number density to the fractional matter density.  In Fourier space, this ratio has been found to depart from the Gaussian expectation by a correction term, $\Delta b(k)$, containing two parts: one that depends on the wavenumber (scale-dependent) and one that does not (scale-independent).  In the limit of small wavenumber, the correction approaches the form $\Delta b(k) \propto \fNL (b_{\mathrm{G}} - 1)  / k^2$, where $b_{\mathrm{G}}$ is the expected Gaussian bias  \citep{2008PhRvD..77l3514D, 2008ApJ...677L..77M,2008PhRvD..78l3507A}.  Based upon this assumed $k^{-2}$ scale-dependence, \citet{2008JCAP...08..031S} have already constrained $\fNL$ to be in the range $-31 <  \fNL < 70$ ($95\%$ limit) using the clustering of massive galaxies and quasars in the Sloan Digital Sky Survey; a result that is competitive with the WMAP7 constraints.  There is naturally a great interest in this method as future large-scale structure surveys of ever-increasing volume utilizing both the galaxy power spectrum and bispectrum may surpass the CMB measurements in constraining $\fNL$ \citep{2004PhRvD..69j3513S,2009ApJ...703.1230J,2010JCAP...07..002N,2011JCAP...04..006B,2012MNRAS.422.2854G}.

Numerical N-body methods have confirmed that the halo bias scales as $k^{-2}$ in the small-$k$ limit in the local-quadratic model, with a redshift dependence inversely proportional to the linear perturbation growth factor, $D(z)$, as predicted by the analytical theory \citep{2008PhRvD..77l3514D,2009MNRAS.396...85D,2010MNRAS.402..191P,2009MNRAS.398..321G,2010JCAP...07..002N,2011JCAP...11..009S}.  However, the amplitude of the bias is still somewhat uncertain, as the N-body simulations so far produce a range of values differing at the $10-20\%$ level \citep{2010AdAst2010E..89D}.  On the other hand, the analytical predictions disagree significantly with results from N-body simulations in models beyond the local-quadratic case \citep{2010PhRvD..81b3006D,2011JCAP...03..017S,2012JCAP...03..002W}.  This motivated \citet{2011PhRvD..84f3512D} to re-examine three analytical derivations of the bias \citep[see also][]{2012JCAP...03..032S}:  1)  An approach based on the statistics of thresholded regions in the density field \citep{2008ApJ...677L..77M}.  2)  A peak-background split (PBS) approach based on the separation of uncorrelated long- and short-wavelength contributions to the \emph{Gaussian} perturbations \citep{2008PhRvD..77l3514D,2008JCAP...08..031S,2010PhRvD..82j3002S}.  3)  A second PBS approach \citep{2010PhRvD..82j3529D} using the conditional halo mass function, which they derive by an extension of the \citet{1974ApJ...187..425P} method to non-Gaussian initial conditions \citep[also see][]{Matarrese:2000pb,Lo-Verde:2008rt}.  This approach is conceptually different from the previous one because it does not involve a separation of scales in the Gaussian perturbations, but instead considers through the conditional mass function how the local halo-abundance depends on the large-scale \emph{non-Gaussian} density contrast.  In addition to showing that the thresholding method cannot be reconciled with N-body simulations, \citet{2011PhRvD..84f3512D} use the last two approaches to derive a new scale-dependent contribution to the bias that was previously overlooked in the literature.  In a companion paper \citep{2011PhRvD..84f1301D}, they showed that the new term is critical for improving the analytical predictions that were previously discrepant with N-body simulations.  

It is important to confirm the above findings with more general methods.  Here, we consider an independent analytical approach to the halo bias - the excursion set formalism \citep{Bond:1991sf,1993MNRAS.262..627L}.  Until recently, the excursion set method had been analytically tractable only in the case with Gaussian initial conditions and, even then, only when a sharp k-space filter was used.  In this case, the dynamics in the excursion-set random-walk model for identifying halos in the initial density field are Markovian.  \citet{2010ApJ...711..907M,2010ApJ...717..515M,2010ApJ...717..526M} recently showed how to extend the excursion set model to include non-Markovian dynamics by formulating it with a path integral.  This breakthrough opens the door to non-Gaussian initial conditions and/or more general filter functions \citep[for a different approach to non-Markovian dynamics, see][]{2012MNRAS.419..132P,2012MNRAS.420.1429P,2012MNRAS.420..369M,2012MNRAS.tmpL.449M,2012arXiv1205.3401M}.  The path-integral approach has been successfully applied in a number of contexts involving PNG \citep{2010ApJ...717..526M,2011JCAP...02..001D,2011MNRAS.412.2587D,2011PhRvD..83b3521D,2011MNRAS.415.1913D,2011MNRAS.418.2403D}.  However, it has not yet been used to derive the scale-dependent correction to the halo bias.  

In what follows, we use the path-integral excursion set method to derive expressions for the conditional collapsed fraction and conditional mass function for non-Gaussian models with general bispectra.  We then use these expressions to compute both scale-dependent and -independent corrections to the halo bias.  In comparison with the results of \citet{2011PhRvD..84f3512D}, our results will come closest to the conclusions of their third approach described above.  In the process of deriving our result, we will be able to investigate more specifically under what circumstances their result is valid.  In particular, the excursion-set approach shows through the non-Markovian dynamics that environmental effects on halo formation can lead to marked differences in halo bias.  We will also quantify these differences here by a perturbative calculation, applied for illustrative purposes to the familiar local-quadratic model, as well as a second example - the so-called orthogonal template.  

The remainder of this paper is organized as follows.  In $\S$ \ref{SEC:statofdensityfield}, we review statistics of the linear density field and define the bispectrum templates used in this work for plotting purposes.  In $\S$ \ref{SEC:PNGpathintegral}, we outline the path-integral excursion set method of \citet{2010ApJ...711..907M,2010ApJ...717..515M,2010ApJ...717..526M}.  In $\S$ \ref{SEC:collapsefraction}, we use the formalism to calculate the conditional collapsed fraction to leading-order, and also compute the next-to-leading order environmental corrections.  We then use the collapsed fraction to obtain expressions for the conditional mass function in $\S$ \ref{SEC:condmassfunc}.  From the conditional mass function, we obtain scale-dependent and -independent linear bias parameters in $\S$ \ref{SEC:halobias}.  Finally, we summarize our results and offer concluding remarks in $\S$ \ref{SEC:conclusion}.  

When plotting our results, we use a fiducial $\Lambda$CDM cosmology with parameters $\Omega_m=0.27$, $\Omega_{\Lambda} = 0.73$, $\Omega_b = 0.046$, $H_0 = 100 h~\mathrm{km~s^{-1}~Mpc^{-1}}$ (with $h=0.7$), $n_s=0.97$ and $\sigma_8=0.82$, consistent with WMAP7 constraints \citep{2011ApJS..192...18K}.  We also employ the linear matter power spectrum of \citet{1999ApJ...511....5E}.


\section{Statistics of density fluctuations in models with primordial non-Gaussianity}
\label{SEC:statofdensityfield}

\subsection{The density contrast and its two- and three-point functions}

The excursion set model is formulated in the Lagrangian picture; it is a method for computing halo statistics from the statistics of the linearly extrapolated initial density fluctuations.  We quantify fluctuations in the linearly extrapolated density field, $\rho_m(\boldsymbol{x},z)$, with the density contrast smoothed on scale $R$ about a point $\boldsymbol{x}$,    

\begin{equation}
\da_R(\mathbf{x},z) = \int{\dd^3 x'~W\left( |\mathbf{x}-\mathbf{x}' |,R\right) \da(\mathbf{x}',z)},
\label{EQ:smootheddelta}
\end{equation}
where $W$ is a spherically symmetric filter function with characteristic scale $R$ and the un-smoothed density contrast is $\delta(\boldsymbol{x},z) \equiv \rho_m(\boldsymbol{x},z)/\bar{\rho}_m(z) - 1$.  The Fourier transform of the smoothed density contrast is related to the primordial Bardeen potential in the matter-dominated epoch through the cosmological Poisson equation\footnote{Strictly speaking, this relation holds only in the synchronous comoving gauge \citep{2012PhRvD..85b3504J}.},

\begin{equation}
\tilde{\da}_R(\boldsymbol{k},z)= \mathcal{M}_R(k,z) \tilde{\Phi}(\boldsymbol{k}).
\end{equation}
Here, we define

\begin{equation}
\mathcal{M}_R(k) \equiv \frac{2}{3} \frac{k^2 T(k) g(z)}{\Omega_m H_0^2 (1+z)} \tilde{W}(k,R),
\end{equation}
where $T(k)$ is the matter transfer function normalized to unity on large scales, $g(z)$ is the linear growth factor of the potential normalized to unity during the epoch of matter domination ($g(0) \approx 0.76$ for our fiducial cosmology), $H_0$ is Hubble's constant, and $\Omega_m$ is the present-day matter density in units of the critical density.  

Gaussian initial density fluctuations are uniquely characterized by their variance,

\begin{equation}
S_R \equiv \sigma^2_R = \langle \da^2_R \rangle_c = \int \frac{\dd^3 \boldsymbol{k}}{(2 \pi)^3} \MM^2_R(k,z) P_{\Phi}(k).
\end{equation}
More generally, higher-order correlation functions are required to characterize non-Gaussian initial fluctuations.  In what follows, we consider non-Gaussianity that is characterized solely through the three-point function, 
 
\begin{align}
\langle \da_{R_1} \da_{R_2} \da_{R_3} \rangle_c  =  & \int \frac{\dd^3 \mathbf{k}_1}{(2 \pi)^3} \frac{\dd^3 \mathbf{k}_2}{(2 \pi)^3} \frac{\dd^3 \mathbf{k}_3}{(2 \pi)^3} \MM_{R_1}(k_1,z) \MM_{R_2}(k_2,z) \MM_{R_3}(k_3,z) & \nonumber \\ &  \times B_{\Phi}(k_1,k_2,k_3) (2 \pi)^3 \da_D(\mathbf{k}_1 + \mathbf{k}_2 + \mathbf{k}_3) \exp\left[-i(\mathbf{k}_1+ \mathbf{k}_2 + \mathbf{k}_3 )\cdot \mathbf{x} \right], &
\end{align}
where $\delta_D$ is the Dirac delta function. We neglect the effects of all higher-order spectra.

The excursion set method is unique among other analytical techniques for treating non-Gaussianities because it depends on $\langle \da_{R_1} \da_{R_2} \da_{R_3}\rangle_c$ for the full range of smoothing scales down to the Lagrangian radius of the halo. As we will see, the mixed three-point correlator $\langle \da_{R_1} \da_{R_1} \da_{R_2} \rangle_c$ will be particularly important in our calculations; it is the source of the scale-dependence of the bias.  We may rewrite this correlator in a way that is convenient for manifesting the scale-dependence by collapsing the delta function, rearranging, and relabeling the integration variables to obtain

\begin{align}
\langle \da_{R_1} \da_{R_1} \da_{R_2}  \rangle_c  =  \int \frac{\dd^3 \mathbf{k}}{(2 \pi)^3} \MM_{R_2}(k,z) P_\phi (k) 4 S_{R_1}  \left[ \frac{1}{ 4 S_{R_1} P_{\phi}(k)}  \int \frac{\dd^3 \mathbf{k}_1}{(2 \pi)^3}   \MM_{R_1}(k_1,z) \MM_{R_1}(q,z)  B_{\Phi}(k,k_1,q) \right],
\end{align}
where $\boldsymbol{q} \equiv  - (\boldsymbol{k_1}+\boldsymbol{k_2})$.  The quantity in the brackets is a form factor denoted by $\mathcal{F}^{(3)}_{R_1}(k)$ in the notation of \citet{2011PhRvD..84f3512D}.  We adopt this notation and express the mixed correlator as

\begin{align}
\langle \da_{R_1} \da_{R_1} \da_{R_2} \rangle_c(\boldsymbol{x)}  =  \int \frac{\dd^3 \mathbf{k}}{(2 \pi)^3}  \MM_{R_2}(k,z) P_\phi (k) 4 S_{R_1}  \mathcal{F}^{(3)}_{R_1}(k) .
\label{EQ:mixedcorrelator}
\end{align}
The advantages of rewriting $\langle \da_{R_1} \da_{R_1} \da_{R_2} \rangle_c$ in this form will become apparent in the following sections. 

\subsection{Phenomenological templates of the primordial bispectrum} 

Primordial bispectra generated by inflationary models vary considerably and can be quite complicated.  For the purpose of computing the effects of a primordial bispectrum on large-scale structure, it is convenient to employ commonly used phenomenological templates.  We will use the local template extensively when plotting our results, where the primordial bispectrum is given by equation (\ref{EQ:localquadraticbispectrum}).  At times, we will find it desirable to bring out the effects of the form factor, $\mathcal{F}^{(3)}_R$, in our final results.  The local template is not ideal for this task because, in this case, $\mathcal{F}^{(3)}_R$ asymptotes to $\fNL$ in the low-$k$ limit.  As a result, we will also use the orthogonal template \citep{2010JCAP...01..028S}, where the bispectrum has the form

\begin{equation}
\begin{split}
B^{\mathrm{ortho}}_{\Phi}(k_1,k_2,k_3) = 6 \fNL^{\mathrm{ortho}} \Biggl[ -3\left(P_{\Phi}(k_1) P_{\Phi}(k_2) + \mathrm{cyc.} \right) - 8\left( P_{\Phi}(k_1) P_{\Phi}(k_2) P_{\Phi}(k_3)  \right)^{2/3} \\ + 3\left( P^{1/3}_{\Phi}(k_1) P^{2/3}_{\Phi}(k_2) P_{\Phi}(k_3) + \mathrm{cyc.}  \right) \Biggr],
\end{split}
\end{equation}
and $\fNL^{\mathrm{ortho}}$ is a constant parameter that determines the amplitude of the bispectrum.  In this case, the form factor scales as $\mathcal{F}^{(3)}_{R} \sim k$ in the low-$k$ regime.

\section{The path-integral generalization of the excursion set formalism to include non-Gaussian initial conditions}
\label{SEC:PNGpathintegral}

Here we summarize the non-Markovian extension of the excursion set formalism by \citet{2010ApJ...711..907M,2010ApJ...717..526M}.  For more details on the original formulation of the excursion set model, we refer the reader to the pioneering paper by \citet{Bond:1991sf} (also see the review of \citet{2007IJMPD..16..763Z} and references therein).  

It is convenient to linearly extrapolate the initial density fluctuations to the present day.  With this choice, the density field stays fixed in time, and therefore so do the variance and three-point function, while the linear over-density threshold for collapse, $\delta_c(z)$, acquires an additional redshift dependence, $\da_c(z) \rightarrow\delta_c(z)/D(z)$, where $D(z)$ is normalized to unity at the present day.   In the excursion set procedure, a filter function with characteristic scale $R$ is centered on a fiducial point in space.  The density contrast is smoothed about that point with some large initial scale, $R_0$, to obtain the smoothed density contrast, $\delta_0$, with corresponding variance $S_0 \equiv \sigma_0^2$.  In what follows, the initial filtering scale always corresponds to the limit, $R_0 \rightarrow \infty$, so that $S_0 =0$ and $\delta_0 =0$.  The scale of the filter function is decreased and the corresponding $\da$ and $S$ are again calculated.  As this procedure is repeated, the stochastic variable $\delta$ executes a random walk.  When $\delta$ \emph{first} exceeds the collapse threshold $\da_c$, the fiducial point is assumed to reside within a halo with mass set by the filter scale.  When numerically evaluating our results, we will use the coordinate-space top-hat filter, where $R$ is assumed to be the initial comoving radius of the collapsed over-density, so that the mass of the halo is to a good approximation given by $M= 4 \pi \bar{\rho}_m R^3/3$.  The Fourier transform of the coordinate-space top-hat filter is

\begin{equation}
\tilde{W}_R(k) = 3 \frac{\sin(kR) - k R \cos(kR)}{(kR)^3}.
\end{equation}
However, for simplicity, and since we are mainly interested in the effects due to \emph{non-Gaussianity}, we will neglect non-Markovian correction terms due to the use of this filter function.  In terms of the derived expressions, this is equivalent to using the sharp $k$-space filter in the formalism. 
 
\citet{2010ApJ...711..907M,2010ApJ...717..526M} showed how to formulate the above model in terms of a path integral.  Consider the ``trajectory" traced out in $(S,\da)$-space.  The starting point is to discretize the ``time" interval $\left[0,S\right]$ so that $S_n = \epsilon n$, where $n \in \{0,1,2,\cdots \}$.  The probability density in the space of trajectories may be written as

\begin{equation}
W(\da_0;\da_1,\ldots,\da_n;S_n) \equiv \langle \da_D(\da(S_1)-\da_1)\ldots \da_D(\da(S_n)-\da_n)\rangle.
\end{equation}
The integral representation of the Dirac delta function,

\begin{equation}
\da_D(\da) = \int_{-\infty}^{\infty}{\frac{\dd \lambda}{2 \pi} e^{-i \lambda \da}},
\end{equation}
is then used to write

\begin{equation}
\label{EQ:Wgen}
W(\da_0;\da_1,\ldots,\da_n;S_n)  = \int{ \mathcal{D} \lambda}  \exp \left( i \sum_{i=1}^n{\lambda_i \da_i} + \sum_{p=2}^{\infty} \frac{(-i)^p}{p!} \sum_{i_1=1}^{n}\ldots \sum_{i_p=1}^{n}  \lambda_{i_1} \ldots \lambda_{i_p} \langle \da(S_{i_1}) \ldots \da(S_{i_p})\rangle_c  \right),
\end{equation}
where we have used the notation,

\begin{equation}
\int\mathcal{D}\lambda \equiv \int_{-\infty}^{\infty}\frac{\dd \lambda_1}{2 \pi} \ldots \frac{\dd \lambda_n}{2 \pi}.
\end{equation}
From here on, we will use the notation $\delta_i \equiv \delta(S_i)$.  The probability density for a trajectory to obtain a value $\delta_n$ at the independent coordinate $S_n$, \emph{without} having crossed $\da_c$, is obtained by integrating $W$ over the space of possible trajectories,
\begin{equation}
\label{EQ:PIgen}
\Pi_{\epsilon}(\da_0;\da_n;S_n) \equiv \int_{-\infty}^{\da_c}{\dd \da_1 \ldots \dd \da_{n-1} W(\da_0;\da_1,\ldots,\da_n;S_n) }.
\end{equation}
Note the dependence of $\Pi_{\epsilon}$ on the connected correlators for scales spanning the ``time" interval.  Through this dependence, the path-integral excursion set formalism contains additional information about the relationship between halo formation and the environment compared to the classic Press-Schechter approach and its simple extensions to the case with PNG. 

The upper limits on the integrals in (\ref{EQ:PIgen}) limit trajectories to stay below the collapse barrier.  In the excursion set model, the fraction of mass contained within halos with masses greater than $M_n$ is equal to the fraction of trajectories that cross $\da_c$ before $S_n(M_n)$.  This fraction is calculated by integrating (\ref{EQ:PIgen}) over all possible $\da_n$ to obtain the fraction of trajectories that reach $S_n$ without having crossed $\delta_c$, and then taking the complement,

\begin{equation}
\fcoll(S_n) = 1-\int_{-\infty}^{\da_c}{\dd\da_n~\Pi_{\epsilon}(\da_0;\da_n;S_n)}.
\label{EQ:fcollgeneral}
\end{equation}
After taking the continuum limit, $\epsilon \rightarrow 0$, the mass function follows from

\begin{equation}
\frac{\dd n}{\dd M}  = \frac{\bar{\rho}_m}{M}  \frac{\dd \fcoll }{\dd S} \left| \frac{\dd S}{\dd M }\right|.
\label{EQ:nmansatz}
\end{equation}

So far the discussion has been completely general.  Let us now consider the evaluation of the above expressions in specific cases of interest.  First, we take the case of Gaussian initial conditions, where the connected correlators with $N\ge3$ vanish, and we are left with
\begin{align}
W(\da_0;\ldots,\da_n;S_n)&  =   \int{\mathcal{D}\lambda ~\exp\left(i \sum_{i=1}^n \lambda_i\da_i - \frac{1}{2} \sum_{i,j=1}^n \lambda_i \lambda_j \langle \da_i \da_j\rangle_c. \right) }.  
\label{EQ:Wg}
\end{align}
The situation simplifies even more if we use the sharp k-space filter function, for which $\langle \da_i \da_j \rangle_c = \mathrm{min}(S_i,S_j)$.  In this case let us define

\begin{equation}
\Wgm (\da_0;\ldots,\da_n;S_n) \equiv \int{\mathcal{D}\lambda ~ \exp\left(i \sum_{i=1}^n \lambda_i\da_i - \frac{1}{2} \sum_{i,j=1}^n \lambda_i \lambda_j \mathrm{min(S_i,S_j)} \right) }.
\label{EQ:Wgm}
\end{equation}
\citet{2010ApJ...711..907M} have shown that the corresponding probability density,
 
 \begin{equation}
\label{EQ:PIGMformal}
\PIgm(\da_0;\da_n;S_n) \equiv \int_{-\infty}^{\da_c}{\dd \da_1 \ldots \dd \da_{n-1}  \Wgm (\da_0;\ldots,\da_n;S_n)},
\end{equation}
yields the standard excursion set result upon taking the continuum limit ($\epsilon \rightarrow 0$), 

\begin{equation}
\label{EQ:PIGM}
\PIgmF(\da_0;\da_n;S_n) = \frac{1}{\sqrt{2 \pi S_n}} \left[ \exp\left( \frac{-\da^2_n}{2 S_n}\right) - \exp\left( - \frac{(2 \da_c-\da_n)^2 }{2 S_n} \right) \right].
\end{equation}
In equations (\ref{EQ:Wgm}), (\ref{EQ:PIGMformal}) and (\ref{EQ:PIGM}), the superscripts ``g" and ``m" represent Gaussian and Markovian respectively;  the latter meaning that the corresponding random walks are Markovian processes - a property exhibited only in the case of Gaussian initial conditions \emph{and} the sharp k-space filter.  \citet{2010ApJ...711..907M} have also shown how to handle non-Markovian effects from the coordinate-space top-hat filter in an approximate way by perturbing $W$ over the Markovian expression, though we do not consider such effects here \citep[see however][]{2011MNRAS.411.2644M}.
  
Let us now turn our attention to the case with PNG - the focus of this work.  As stated in the last section, we neglect the effects of the higher-order connected N-point functions, where $N>3$ (i.e. we consider only the primordial bispectrum from PNG in our calculations) so that

\begin{align}
W(\da_0;\ldots,\da_n;S_n)&  =   \int{\mathcal{D}\lambda ~\exp\left(i \sum_{i=1}^n \lambda_i\da_i - \frac{1}{2} \sum_{i,j=1}^n \lambda_i \lambda_j \langle \da_i \da_j\rangle_c \right)\times \exp \left( \frac{(-i)^3}{6} \sum_{i,j,k=1}^n \lambda_i \lambda_j \lambda_k~\coco \right) }.
\label{EQ:WPNG}
\end{align}
From here on we will use the notation $\pd_i \equiv \pd / \pd \da_i$.  Expanding the second exponential in (\ref{EQ:WPNG}) under the assumption of small $\langle \da_i \da_j \da_k \rangle_c$ and using the property, $\lambda \exp( i \lambda \da_i) = -i\pd_i \exp(i \lambda \da_i )$, equation ({\ref{EQ:WPNG}}) becomes

\begin{equation}
 W(\da_0;\ldots,\da_n;S_n) \approx  W^{\mathrm{gm}}(\da_0;\ldots,\da_n;S_n)  - \frac{1}{6}  \sum_{i,j,k=1}^{n}{\coco~\pd_i \pd_j \pd_k W^{\mathrm{gm}}(\da_0;\ldots,\da_n;S_n)}.
\label{EQ:WPNGexpansion}
\end{equation}
This expansion\footnote{\citet{2011JCAP...02..001D} point out in their mass function calculation that the second term on the right-hand side of equation (\ref{EQ:WPNGexpansion}) yields a term that goes as $\delta^3_c \langle \da^3 \rangle_c /\sigma^6$, which can be of order of unity for very large masses (see their Figure 2), indicating that the expansion breaks down in that regime. } \citep{2010ApJ...717..526M} forms the backbone of our perturbative calculation of the collapsed fraction described in the next section.

     
\section{The conditional collapsed fraction}
\label{SEC:collapsefraction}
 
Consider a large, spherical region of the Universe with Lagrangian radius $R=R_l$, containing mass $M_l$, and with a smoothed linear density contrast, $\delta_l$.  The conditional collapsed fraction, $\fcoll(S_s|\da_l,S_l)$, is the fraction of mass in halos with masses between some lower threshold, $M_s$ (corresponding to smoothing scale $R=R_s$), and $M_l$.  It is derived within the excursion set formalism from the conditional probability for a trajectory passing through $(S_l,\da_l)$ to reach $(S_s,\da_s)$, without ever having crossed the barrier $\delta_c$.  This probability density is represented in discrete form by \citep{2011MNRAS.411.2644M,2011MNRAS.415.1913D,2011MNRAS.418.2403D}
\begin{equation}
P(\da_s,S_s|\da_l,S_l)  =  \frac{P(\da_l,S_l,\da_s,S_s)}{P(\da_l,S_l)} = \frac{\int_{-\infty}^{\da_c}{\dd\da_1 \ldots \dd\da_{l-1}} \int_{-\infty}^{\da_c}{\dd\da_{l+1}\ldots\dd\da_{s-1} W\left(\da_0;\da_1,\ldots,\da_s;S_s \right)}}{ \int_{-\infty}^{\da_c}{\dd\da_1 \ldots \dd\da_{l-1}  W\left(\da_0;\da_1,\ldots,\da_l;S_l \right) }}.
\label{EQ:cond_P}
\end{equation}
On the right hand side of (\ref{EQ:cond_P}), ``$s$" and ``$l$" are integers given by $s=S_s/\epsilon$ and $l=S_l/\epsilon$ respectively and $l<s$.  The conditional collapsed fraction corresponds to the fraction of excursion-set trajectories that first cross $\delta_c$ between the scales $S_l$ and $S_s$, which can be obtained by integrating equation (\ref{EQ:cond_P}) over $\dd \delta_s$ and taking its complement,
\begin{equation}
\fcoll(S_s|\da_l,S_l) = 1-\int_{-\infty}^{\da_c}{\dd\da_s~P(\da_s,S_s|\da_l,S_l)}.
\label{EQ:cond_cu_F}
\end{equation}
We start the evaluation of equation (\ref{EQ:cond_P}) by expanding $W(\da_0;\ldots,\da_s;S_s)$ in the numerator and keeping terms up to first order in the three-point correlator,

\begin{align}
W(\da_0;\ldots,\da_s;S_s)&  =   \int{\mathcal{D}\lambda  \exp\left\{i \sum_{i=1}^{s} \lambda_i\da_i - \frac{1}{2}\sum_{i,j=1}^{s} \lambda_i \lambda_j \langle \da_i \da_j\rangle_c + \frac{(-i)^3}{6} \sum_{i,j,k=1}^{s}\lambda_i \lambda_j \lambda_k~\coco \right\}  } \nonumber \\  & \approx  W^{\mathrm{gm}}(\da_0;\ldots,\da_s;S_s)  - \frac{1}{6}  \sum_{i,j,k=1}^{s}{\coco~\pd_i \pd_j \pd_k W^{\mathrm{gm}}(\da_0;\ldots,\da_s;S_s)}, 
\label{EQ:W}
\end{align}
where the first terms on the right-hand side correspond to the Gaussian contribution and the non-Gaussian correction respectively.  Similarly, in the denominator of (\ref{EQ:cond_P}), we have

\begin{align}
W(\da_0;\ldots,\da_l;S_l)&  \approx  W^{\mathrm{gm}}(\da_0;\ldots,\da_l;S_l)  - \frac{1}{6}  \sum_{i,j,k=1}^{l}{\coco~\pd_i \pd_j \pd_k W^{\mathrm{gm}}(\da_0;\ldots,\da_l;S_l)}.
\label{EQ:Wdenominator}
\end{align}

\begin{figure}
\begin{center}
\resizebox{8.5cm}{!}{\includegraphics{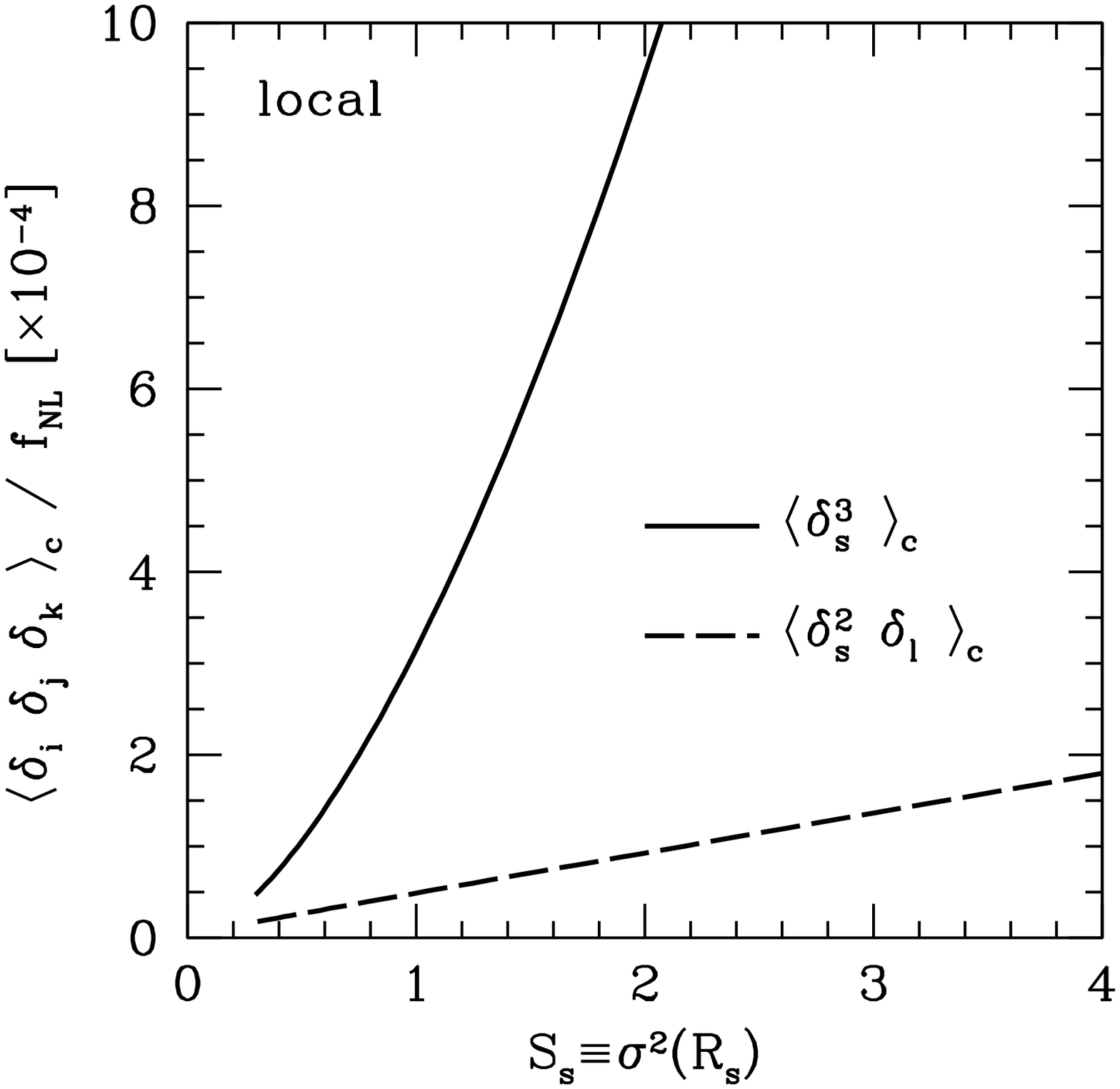}} \hspace{0.13cm}
\resizebox{8.5cm}{!}{\includegraphics{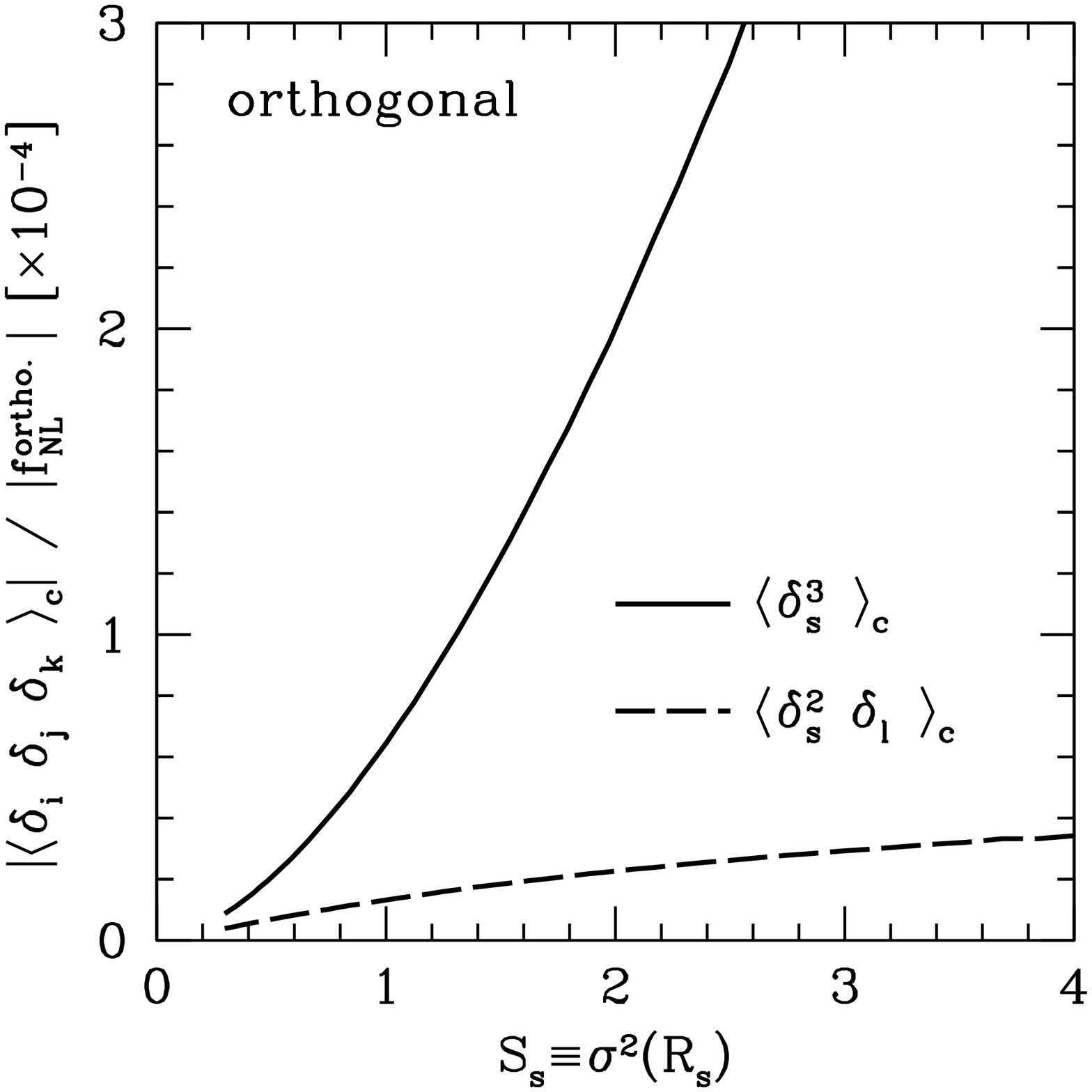}}
\end{center}
\caption{The three-point connected correlators, $\langle \da_s^3 \rangle_c$ and $\langle \da_s^2 \da_l \rangle_c$, which appear in the leading-order conditional collapsed fraction for the local (left) and orthogonal (right) templates.  We assume a fixed $M_l = 10^{16} \Msun/h$ ($R_l=45.3$ Mpc), corresponding to $S_l=0.07$.    }
\label{FIG:threept}
 \end{figure}

In the numerator, we perform a separation between the density contrast smoothed on the halo ($R_s$) and environmental ($R_l$) scales.  This is achieved by breaking up the sum in the second term of (\ref{EQ:W}),

\begin{align}
\label{EQ:sumbreak}
 \sum_{i,j,k=1}^{s}{\coco~\pd_i \pd_j \pd_k} = &  \sum_{i,j,k=1}^{l}{\coco~\pd_i \pd_j \pd_k}  +  3  \sum_{i,j=1}^{l}\sum_{k=l+1}^{s}{\coco~\pd_i \pd_j \pd_k} \\  & + 3  \sum_{i=1}^{l}\sum_{j,k=l+1}^{s}{\coco~\pd_i \pd_j \pd_k}+\sum_{i,j,k=l+1}^{s}{\coco~\pd_i \pd_j \pd_k}. \nonumber
\end{align}
The remaining details of the collapsed fraction calculation are reserved for the appendix due to their technical nature.  An important detail, however, is that the triple sums in equation (\ref{EQ:sumbreak}) become triple integrals over $S_i$, $S_j$, and $S_k$ in the continuum limit, which are not analytically tractable.  However, we can move forward analytically with an extension of the approach of \citet{2010ApJ...717..526M} (also see \cite{2011MNRAS.415.1913D} and \citet{2011MNRAS.418.2403D} ).  We employ a Taylor expansion of $\langle \da_i \da_j \da_k \rangle_c$ about the point $(S_a,S_b,S_c)$,    

\begin{equation}
\langle \da_{i} \da_{j} \da_{k} \rangle_c = \sum_{p,q,r=0}^{\infty} \frac{(-1)^{p+q+r}}{p! q! r!} (S_a-S_i)^p (S_b-S_j)^q (S_c-S_k)^r \left[ \frac{\pd^p}{\pd S_i^p} \frac{\pd^q}{\pd S_j^q} \frac{\pd^r}{\pd S_k^r }  \langle \da_i \da_j \da_k \rangle_c \right]_{S_i = S_a, S_j = S_b, S_k = S_c}.
\label{EQ:sysexpansion}
\end{equation}
For illustrative purposes, in Figure \ref{FIG:threept} we show the three-point connected correlators, $\langle \da_s^3 \rangle_c$ and $\langle \da_s^2 \da_l \rangle_c$, for $M_l = 10^{16}\Msun/h$, corresponding to $S_l=0.07$ in our fiducial cosmology.  In the remainder of this work we will consider the terms in (\ref{EQ:sysexpansion}) originating from the $p=q=r=0$ and $p+q+r=1$ contributions.  In keeping with the terminology of \citet{2010ApJ...717..526M}, we will refer to the $p=q=r=0$ terms as ``leading-order," while the $p+q+r=1$ terms will be called ``next-to-leading order."  We again emphasize that all results in this paper are to first order in the three-point correlator.    

\subsection{The leading-order terms}

To obtain the leading-order terms in equation (\ref{EQ:sumbreak}), we expand each of the correlators about the endpoints of the sum in which they appear, and keep only the constant terms with $p=q=r=0$, so that (\ref{EQ:sumbreak}) becomes
  
\begin{align}
\label{EQ:simplifiedsum}
 \sum_{i,j,k=1}^{s}{\coco~\pd_i \pd_j \pd_k} \approx ~ & \langle \da_l^3 \rangle_c \sum_{i,j,k=1}^{l}\pd_i \pd_j \pd_k +  3 \langle \da_s \da_l^2  \rangle_c  \sum_{i,j=1}^{l}\sum_{k=l+1}^{s}{\pd_i \pd_j \pd_k} \\  & + 3  \langle \da_s^2 \da_l \rangle_c \sum_{i=1}^{l}\sum_{j,k=l+1}^{s}{\pd_i \pd_j \pd_k}+ \langle \da_s^3 \rangle_c \sum_{i,j,k=l+1}^{s}{\pd_i \pd_j \pd_k}. \nonumber
\end{align}
The remaining steps of the calculation are detailed in Appendix {\ref{APP:leadingorder}}.  The final expression for the conditional collapsed fraction up to leading order is

\begin{equation}
\fcoll=\fcollgm+f^{(1)}_{\mathrm{coll}},
\label{EQ:fcollform}
\end{equation}
where
\begin{align}
f^{(1)}_{\mathrm{coll}} = & \frac{\Afd}{3}  \left(  \frac{\delta_c-\delta_l}{S_s-S_l}   - \frac{1}{\da_c-\da_l} \right) \frac{\pd \fcollgm}{\pd S_s} + \frac{\Bfd}{S_l} \left[ \da_c - ( \da_c -\da_l) \coth \left( \frac{\da_c^2-\da_c \da_l }{ S_l} \right)  \right] \frac{\pd \fcollgm}{\pd S_s}  \nonumber \\ & + \Cfd \cdot \frac{S_s-S_l}{S_l^2 (\da_c - \da_l)} \left\{ \da_l^2- S_l  - 2 (\da_c^2-\da_c\da_l) \left[\coth \left( \frac{ \da_c^2-\da_c \da_l }{ S_l} \right)  - 1 \right] \right\} \frac{\pd \fcollgm}{\pd S_s},
\label{EQ:fcoll}
\end{align}
and the connected correlators evaluated at the two scales enter through

\begin{align}
\label{EQ:Afunc}
& \mathcal{A} \equiv \mathcal{A}(S_l,S_s) = \langle \da^3_s \rangle_c -  \langle \da^3_l \rangle_c + 3~\langle \da^2_l \da_s \rangle_c - 3~\langle \da_l \da^2_s \rangle_c \\ \nonumber \\
\label{EQ:Bfunc}
& \mathcal{B} \equiv \mathcal{B}(S_l,S_s)  =  \langle \da^3_l \rangle_c +  \langle \da_l \da^2_s\rangle_c - 2~\langle \da^2_l \da_s \rangle_c  \\ \nonumber \\
\label{EQ:Cfunc}
& \mathcal{C} \equiv \mathcal{C}(S_l,S_s)  =  \langle \da^2_l \da_s \rangle_c -  \langle \da^3_l \rangle_c.
\end{align}
Here, the super-script in the second term on the right-hand side of equation (\ref{EQ:fcollform}) denotes the term that is leading-order in the expansion (\ref{EQ:sysexpansion}), while

\begin{equation}
\fcollgm = \mathrm{erfc}\left[ \frac{\delta_c - \delta_l}{\sqrt{2 [S_s - S_l]}} \right] 
\end{equation}
and

\begin{equation}
 \frac{\pd \fcollgm}{\pd S_s} = \frac{(\da_c - \da_l)}{\sqrt{2 \pi} (S_s - S_l)^{3/2}}  \exp\left[ -\frac{(\da_c - \da_l)^2}{2 (S_s - S_l)} \right],
\end{equation}
are the standard excursion set expressions for the conditional collapsed fraction and first-crossing rate in the case of Gaussian initial conditions and Markovian random walks. Note that  (\ref{EQ:fcollform}) takes the form of the Gaussian and Markovian result plus correction terms proportional to the three-point connected correlators.

\begin{figure}
\begin{center}
\resizebox{8.5cm}{!}{\includegraphics{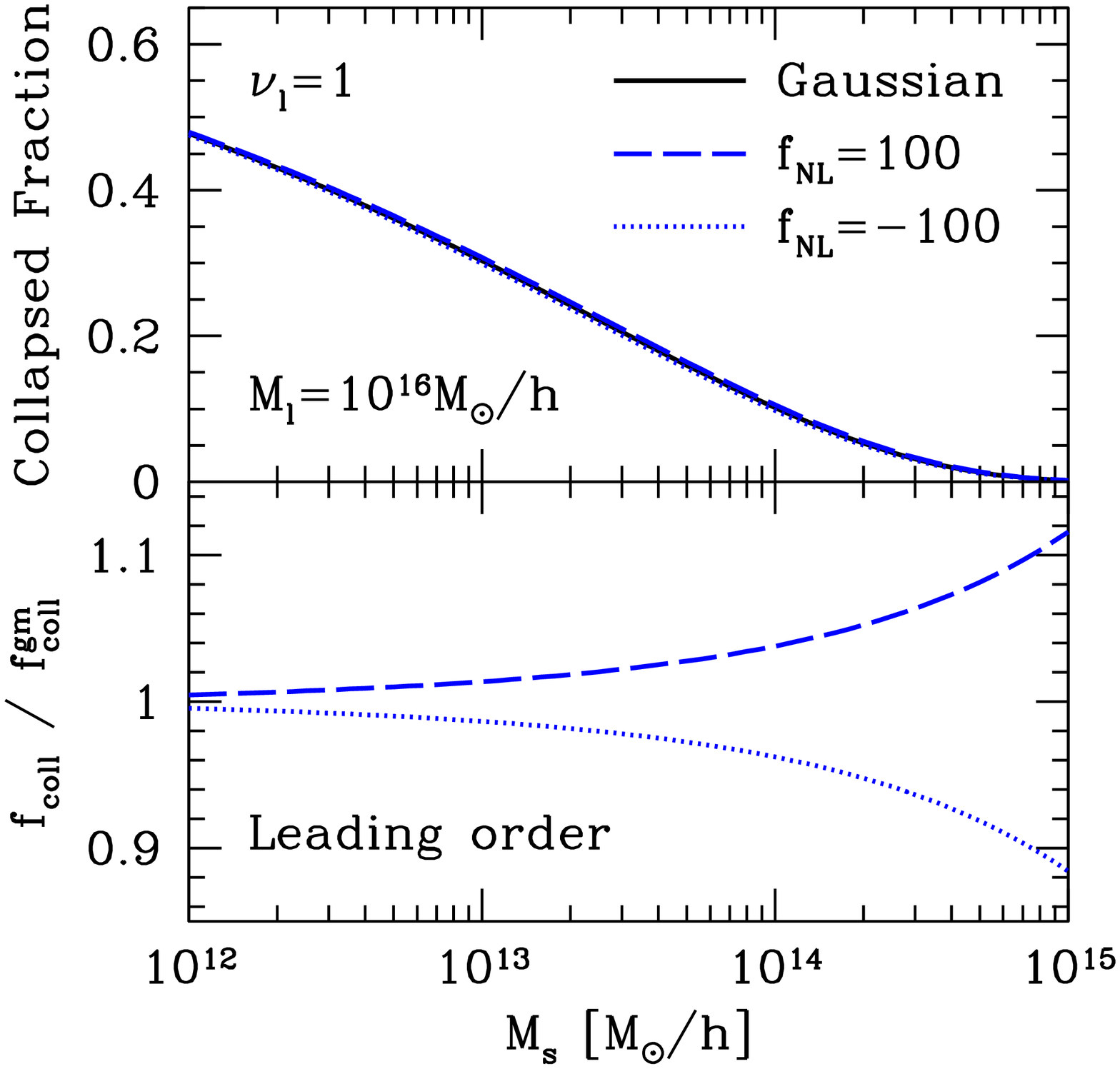}}\hspace{0.25cm}
\resizebox{8.5cm}{!}{\includegraphics{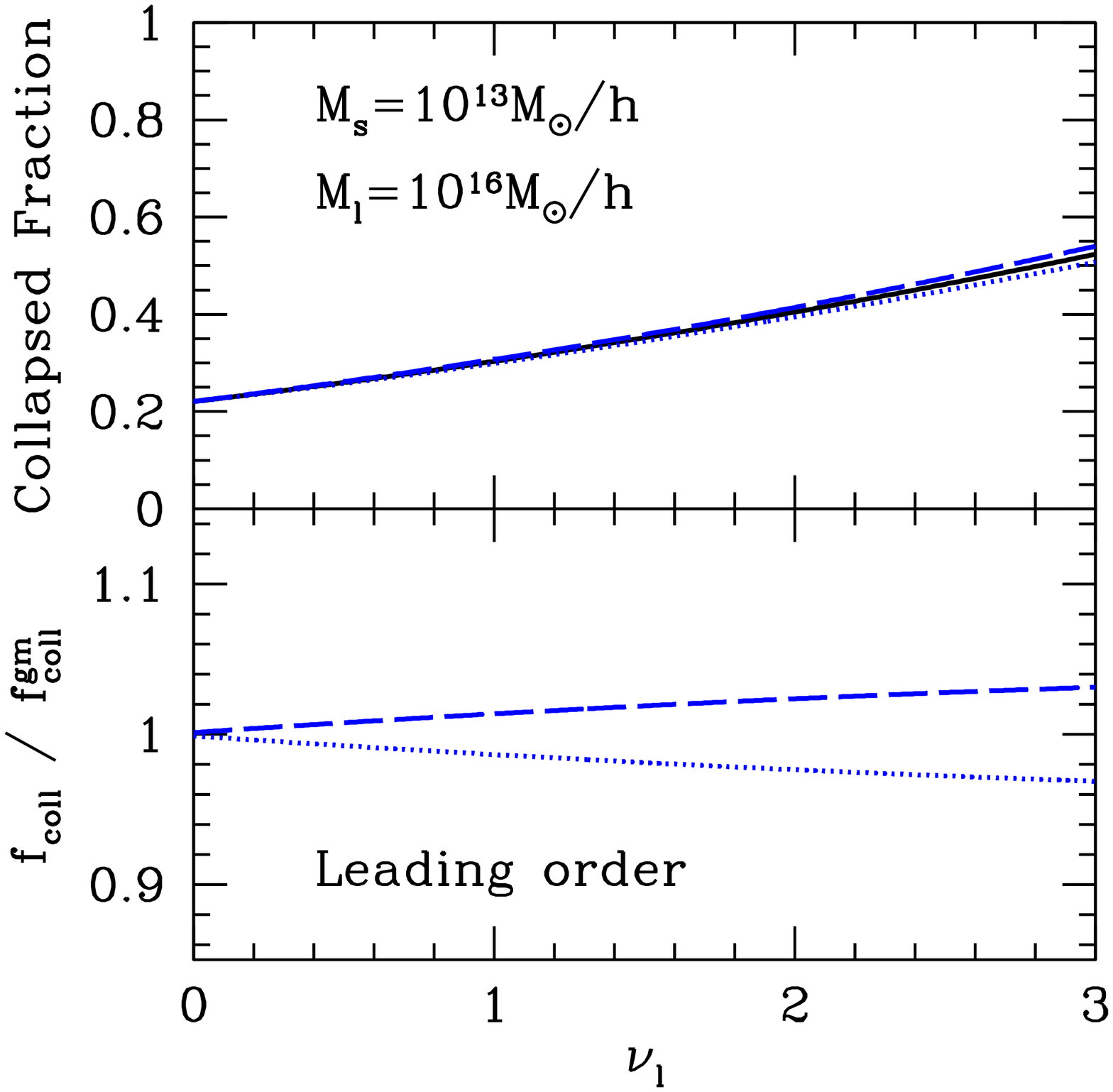}}
\end{center}
\caption{Left panel: the conditional collapsed fraction up to leading-order at $z=0$ for the local template of PNG as a function the lower mass-threshold, $M_s$, for fixed environmental mass $M_l$.  We assume an environmental density fluctuation of mass $M_l=10^{16}\Msun/h$ ($R_l=45.3$ Mpc), with peak-height $\nu_l = \delta_l/\sqrt{S_l} = 1$.  The bottom panel shows its ratio to the Gaussian case.  Right panel: same as the left panel, but as a function of the environmental peak-height, for fixed $M_s=10^{13}\Msun/h$.    }
\label{FIG:fcoll_1storder}
 \end{figure}

Figure \ref{FIG:fcoll_1storder} shows the conditional collapsed fraction at $z=0$ to leading order for the local template.  The top panel on the left shows $\fcoll$ as a function of $M_s$ for a typical large-scale mass fluctuation with $M_l=10^{16}\Msun/h$.  Here, what we mean by a ``typical" fluctuation is that the environmental peak height, $\nu_l \equiv \delta_l / \sqrt{S_l}$ is unity.   For $M_l=10^{16}\Msun/h$, this translates to $\delta_l =0.27$ in our fiducial cosmology.  PNG has a larger impact on the conditional collapsed fraction when only larger halos are included in the latter (case with larger $M_s$).  PNG has a smaller effect when the conditional collapsed fraction is dominated by less massive halos (case with smaller $M_s$), whose abundances are not as strongly affected by PNG.  The plot on the right of Figure \ref{FIG:fcoll_1storder} shows $\fcoll$ for a fixed $M_s=10^{13}\Msun/h$ and $M_l=10^{16}\Msun/h$ as a function of the environmental peak height, $\nu_l$.  It shows that PNG has a larger impact on the conditional collapsed fraction within higher significance (i.e. rarer) peaks.

\subsection{The next-to-leading order terms}

\begin{figure}
\begin{center}
\resizebox{9.5cm}{!}{\includegraphics{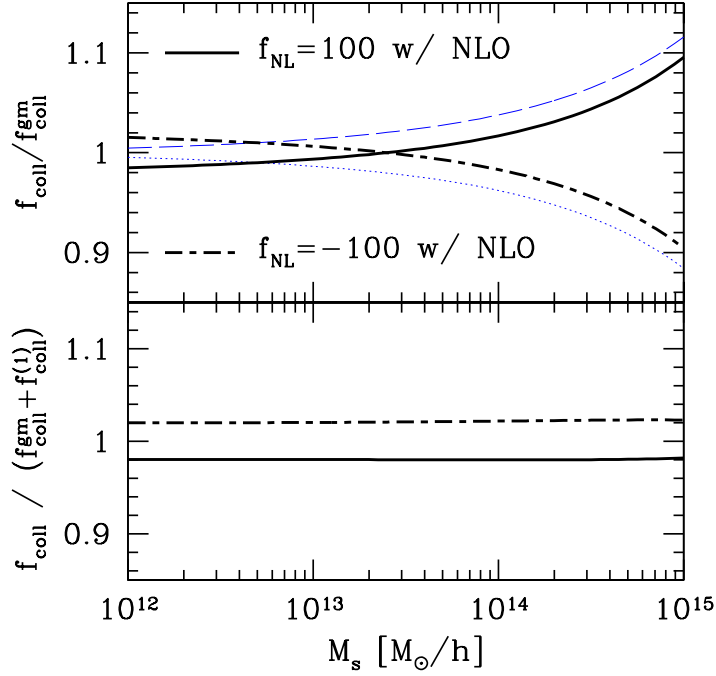}}
\end{center}
\caption{The effects of the next-to-leading order (NLO) corrections to the conditional collapsed fraction.  All parameters are the same as the left panel of Figure \ref{FIG:fcoll_1storder}.  The top panel shows the ratios of both the leading-order (thin, dashed and dotted) and next-to-leading order (thick, solid and dot-dashed) cases to the Gaussian and Markovian case, while the bottom panel shows the fractional change from adding the next-to-leading order corrections.}
\label{FIG:fcoll_2ndorder}
 \end{figure}

The next-to-leading order terms in the conditional collapsed fraction are obtained from the $p+q+r=1$ terms in equation (\ref{EQ:sysexpansion}).  We adopt a short-hand notation similar to \citet{2010ApJ...717..526M} for the derivatives of the three-point correlators,

\begin{equation}
G^{(p,q,r)}_{a,b,c} \equiv \left. \frac{\pd^p}{\pd S_i^p} \frac{\pd^q}{\pd S_j^q} \frac{\pd^r}{\pd S_k^r }  \langle \da_i \da_j \da_k \rangle_c \right|_{S_i = S_a, S_j = S_b, S_k = S_c}.
\end{equation}
The details of the calculation are given in Appendix \ref{APP:nextoleading}.  The next-to-leading order terms in the conditional collapsed fraction are

\begin{equation}
\begin{split}
\label{EQ:f_coll_2}
\fcoll^{(2)} &=\left[ 2 (G^{(1,0,0)}_{l,l,l} -G^{(1,0,0)}_{l,l,s})  \frac{(\dcl)(2\dc-\dl)}{S_l} -    (G^{(1,0,0)}_{l,l,l} -2 G^{(1,0,0)}_{l,l,s})\frac{(\dcl)^2}{\ssl} \right.\\ 
&-\left.G^{(1,0,0)}_{l,s,s}\frac{(\dcl)^2}{\ssl}-2G^{(0,0,1)}_{l,l,s}\frac{\ssl}{S_l^2}\da_c(\dcl) \right]    \left[1-\coth\left(\frac{\dc^2-\dc\dl}{S_l}\right)\right]  \frac{(S_s-S_l)}{(\da_c-\da_l)} \frac{\pd \fcollgm}{\pd S_s}\\ 
&+\frac{1}{S_l}\left[\left(G^{(0,0,1)}_{l,l,s}-G^{(0,1,0)}_{l,s,s}\right)(\dcl)+ G^{(0,0,1)}_{l,l,s} \frac{\da_c(\dcl)^2}{S_l}\right]  \left[1-\coth\left(\frac{\dc^2-\dc\dl}{S_l}\right)\right] \fcollgm \\ 
&+\left[\frac{2 G^{(0,1,0)}_{l,s,s}- G^{(0,0,1)}_{l,l,s}-G^{(1,0,0)}_{s,s,s}}{\sqrt{2\pi(\ssl)}}+\frac{G^{(0,0,1)}_{l,l,s}}{S_l}\sqrt{\frac{\ssl}{2\pi}}\left(1-\frac{\dl^2}{S_l}\right)\right]   \frac{\sqrt{2 \pi} (S_s-S_l)^{3/2}}{(\da_c-\da_l)} \frac{\pd \fcollgm}{ \pd S_s}  \\ 
&-\left[G^{(0,1,0)}_{l,s,s}\frac{\dl}{S_l}-\frac{G^{(0,0,1)}_{l,l,s}}{2S_l}\left(3\dl-\dc+\frac{(\dcl)\dl^2}{S_l}\right)\right] \fcollgm.
\end{split}
\end{equation}
Figure \ref{FIG:fcoll_2ndorder} shows the effect of adding the next-to-leading order terms in the local template.  As in Figure \ref{FIG:fcoll_1storder}, the environmental mass and peak height are set to $M_l=10^{16}\Msun/h$ and $\nu_l=1$ respectively.   The top panel shows the ratios of both $\fcollgm+\fcoll^{(1)}$ (thin, dashed and dotted) and $\fcollgm+ \fcoll^{(1)}+\fcoll^{(2)}$ (thick, solid and dot-dashed) to the Gaussian and Markovian collapsed fraction, $\fcollgm$.  The bottom panel shows the fractional change from adding $\fcoll^{(2)}$, indicating in this example that the next-to-leading order corrections act to suppress the leading-order result by $\sim 2 \%$ for $\fNL=100$ and to enhance by the same factor for $\fNL=-100$.

\section{The Conditional mass function}
\label{SEC:condmassfunc}

\begin{figure}
\begin{center}
\resizebox{9.5cm}{!}{\includegraphics{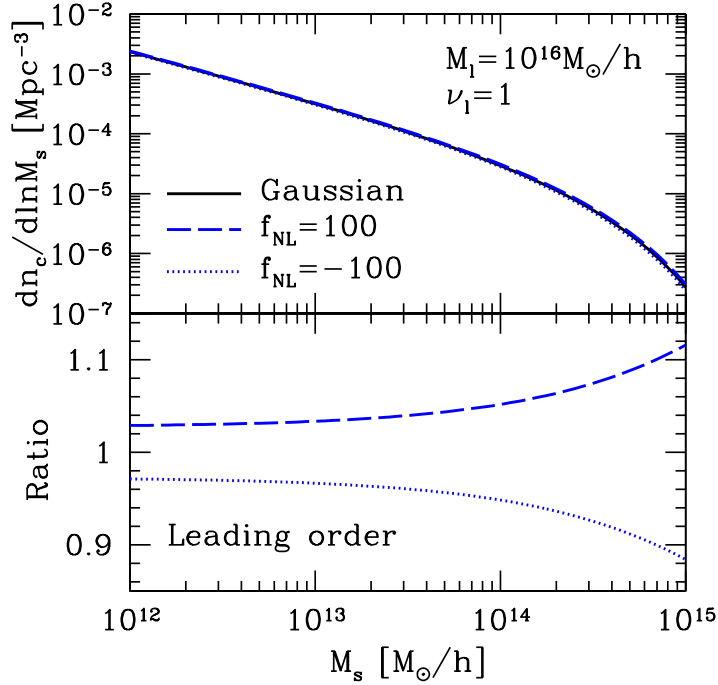}}
\end{center}
\caption{The conditional mass function up to leading-order at $z=0$ for the local template of PNG.  The bottom panel shows the ratio of the non-Gaussian cases to the Gaussian case.}
\label{FIG:nm_cond}
 \end{figure}

The conditional mass function is an important ingredient for deriving the halo bias in the next section.  We therefore devote this section towards obtaining leading-order and next-to-leading order expressions for it.  The conditional mass function may be written in terms of the first-crossing rate,

\begin{equation}
\frac{\dd n_c}{\dd m} = \frac{\bar{\rho}_m}{m} \frac{\pd \fcoll}{\pd S_s} \left| \frac{\dd S_s}{\dd m} \right|.
\label{EQ:nm_cond_ansatz}
\end{equation}
The middle factor on the right-hand side requires taking a derivative with respect to $S_s$ of the collapsed fraction that we obtained in the last section.  Let us now consider the leading-order contribution, which may be written as

\begin{equation}
\frac{\pd \fcoll^{(1)}}{\pd S_s} = \frac{\pd \fcollgm}{\pd S_s} \left[ \left( \frac{\pd \Afd }{\pd S_s} + \alpha~\Afd\right) \chi+  \left( \frac{\pd \Bfd }{\pd S_s} + \beta~\Bfd\right) \psi +  \left( \frac{\pd \Cfd }{\pd S_s} + \gamma~\Cfd\right) \omega \right],
\label{EQ:dfcolldS}
\end{equation}
where we define the following auxiliary functions to compactly express the conditional mass function, as well as greatly simplify the bias calculation below:

\begin{equation}
\alpha \equiv \frac{(\da_c - \da_l)^2}{2 (S_s-S_l )^2} - \frac{5}{2(S_s - S_l)} - \frac{1}{(\da_c-\da_l)^2 - (S_s-S_l)}
\end{equation}

\begin{equation}
 \beta \equiv \frac{(\da_c-\da_l)^2-3(S_s-S_l)}{2(S_s-S_l)^2}
\end{equation}

\begin{equation}
 \gamma \equiv \frac{(\da_c-\da_l)^2-(S_s-S_l)}{2(S_s-S_l)^2}
\end{equation}

\begin{equation}
 \chi  \equiv \frac{1}{3}  \left\{  \frac{\delta_c-\delta_l}{S_s-S_l} - \frac{1}{\da_c-\da_l} \right\}
\end{equation}

\begin{equation}
  \psi \equiv \frac{1}{S_l} \left[ \da_c - ( \da_c -\da_l) \coth \left( \frac{\da_c^2-\da_c \da_l }{ S_l} \right)  \right]
\end{equation}

\begin{equation}
 \omega \equiv \frac{S_s-S_l}{S_l^2 (\da_c - \da_l)} \left\{ \da_l^2- S_l  - 2 (\da_c^2-\da_c\da_l) \left[\coth \left( \frac{ \da_c^2-\da_c \da_l }{ S_l} \right)  - 1 \right] \right\}.
\end{equation}
At first glance, such auxiliary definitions might seem cumbersome, but the payoff is apparent for those who follow the details of the bias calculation given in the appendix.  Using this notation, the conditional mass function up to leading-order is
\begin{equation}
\frac{\dd n_c}{\dd m} = \left( \frac{\dd n_c}{\dd m} \right)^{\mathrm{PS}} \left[ 1 + \left( \frac{\pd \Afd }{\pd S_s} + \alpha~\Afd\right) \chi+  \left( \frac{\pd \Bfd }{\pd S_s} + \beta~\Bfd\right) \psi +  \left( \frac{\pd \Cfd }{\pd S_s} + \gamma~\Cfd\right) \omega \right],
\label{EQ:nm_cond}
\end{equation}
where 

\begin{equation}
\left( \frac{\dd n_c}{\dd m} \right)^{\mathrm{PS}} = \frac{\bar{\rho}_m}{m} \frac{\pd \fcollgm}{ \pd S_s} \left| \frac{\dd S_s}{ \dd m} \right| =  \frac{\rho_m}{m} \frac{(\da_c-\da_l)}{\sqrt{2 \pi} (S_s-S_l)^{3/2}}  \exp \left[-\frac{(\da_c-\da_l)^2}{2 (S_s - S_l)}  \right] \left| \frac{\dd S_s}{ \dd  m} \right|
\end{equation}
is the conditional Press-Schechter mass function \citep{1993MNRAS.262..627L}.  Aside from our restriction to the spherical collapse model, we note that our expression differs from the moving-barrier results of \citet{2011MNRAS.418.2403D}, since we include terms involving $\langle \da_s \da_l^2 \rangle_c$ and $\langle \da_l^3 \rangle_c$, which are important unless $S_l$ is very small.  The upper panel of Figure \ref{FIG:nm_cond} shows equation (\ref{EQ:nm_cond}) at $z=0$ for fixed $M_l=10^{16}\Msun/h$, $\nu_l = 1$, and $\fNL=\pm100$, while the bottom panel shows their ratio to the Gaussian case.  Before moving on to next-to-leading order contributions, we note that the unconditional non-Gaussian mass function can be obtained from equation (\ref{EQ:nm_cond}) by taking the limit $\da_l \rightarrow 0$ and $S_l \rightarrow 0$.  In this limit, equation (\ref{EQ:nm_cond}) recovers the mass function originally derived by \citet{Lo-Verde:2008rt} using the Edgeworth expansion and the Press-Schechter approach (i.e. integrating the probability density function for $\da$), though the apparent correspondence of their result with the excursion set result is not trivial\footnote{See \citet{2010ApJ...717..526M} for a detailed comparison of the Press-Schechter and excursion set approaches in the context of PNG. }  

A full calculation of the next-to-leading order terms would require differentiating equation (\ref{EQ:f_coll_2}) with respect to $S_s$.  Fortunately, we do not need to perform this cumbersome task for the bias calculation below.  There, we may focus on the large-scale environment regime, which is much simpler in practice because many of the above terms disappear in that limit.  For this reason, we do not consider the full next-to-leading order corrections to the conditional mass function here, but instead consider the case where $|\delta_l| \ll 1$ and $S_l \ll 1$.  In this regime, we keep only terms in the leading-order expression involving correlators that are linear in $\delta_l$, so that $\Afd \rightarrow \langle \da_s^3 \rangle_c - 3 \langle \da_l \da_s^2\rangle_c$, $\Bfd \rightarrow \langle \da_l \da_s^2 \rangle_c$, and $\Cfd \rightarrow 0$.  The next-to-leading order corrections to the first-crossing rate immediately simplify because of the $1-\coth\left[ (\da_c^2 - \da_c \da_l ) / S_l \right]$ factors appearing in (\ref{EQ:f_coll_2}), which are zero in the limit of $S_l \rightarrow 0$, and suppress any terms they multiply.  Of the remaining terms, we neglect those involving $G^{(0,0,1)}_{l,l,s}/S_l$ since they are several orders of magnitude smaller than the $G^{(1,0,0)}_{s,s,s}$ and $G^{(0,1,0)}_{l,s,s}$ terms when  $|\delta_l| \ll 1$ and $S_l \ll 1$.  With these simplifications, the conditional first-crossing rate up to next-to-leading order in the large-scale environment regime ($|\delta_l| \ll 1$, $S_l \ll 1$) is

\begin{equation}
\frac{\pd \fcoll}{\pd S_s} = \frac{\pd \fcollgm}{\pd S_s}+\frac{\pd \fcoll^{(1)}}{\pd S_s} + \frac{\pd \fcoll^{(2)}}{\pd S_s}, 
\label{EQ:firstcrossingsum}
\end{equation}
where

\begin{equation}
\frac{\pd \fcoll^{(1)}}{\pd S_s} \approx \frac{\pd \fcollgm}{\pd S_s} \left[ \left( \frac{\pd \langle \da_s^3 \rangle_c }{\pd S_s} - 3 \frac{\pd \langle \da_s^2 \da_l \rangle_c }{\pd S_s} + \alpha \langle \da_s^3 \rangle_c - 3 \alpha \langle \da_s^2 \da_l \rangle_c \right) \chi+  \left( \frac{\pd \langle \da_s^2 \da_l \rangle_c }{\pd S_s} + \beta~\langle \da_s^2 \da_l \rangle_c\right) \psi  \right],
\label{EQ:dfcolldS_1storder}
\end{equation}
and

\begin{equation}
\begin{split}
\frac{\pd \fcoll^{(2)}}{\pd S_s} \approx  \left(  2 G^{(0,1,0)}_{l,s,s} - G^{(1,0,0)}_{s,s,s}  \right) \frac{3 \chi}{2} \frac{\pd \fcollgm}{\pd S_s}   + \left( 2 \frac{ \pd G^{(0,1,0)}_{l,s,s}}{\pd S_s} -  \frac{ \pd G^{(1,0,0)}_{s,s,s}}{\pd S_s}  \right)  \mu \frac{ \pd \fcollgm}{\pd S_s}  
- G^{(0,1,0)}_{l,s,s} \frac{\da_l}{S_l} \frac{\pd \fcollgm}{\pd S_S} - \frac{\pd G^{(0,1,0)}_{l,s,s}}{\pd S_s} \frac{\da_l}{S_l} \fcollgm.
\end{split}
\label{EQ:dfcolldS_2lsl}
\end{equation}
For convenience later on, we have defined another auxiliary function,

\begin{equation}
\mu \equiv \frac{S_s -S_l}{\da_c - \da_l},
\end{equation}
and have used the fact that 

\begin{equation}
 \frac{\pd }{\pd S_s} \left( \mu \frac{ \pd \fcollgm}{\pd S_s} \right)  = \frac{3 \chi}{2} \frac{\pd \fcollgm}{\pd S_s} 
\end{equation}
to obtain equation (\ref{EQ:dfcolldS_2lsl}).  The next-to-leading order conditional mass function in the large-scale environment limit then follows from (\ref{EQ:nm_cond_ansatz}) and (\ref{EQ:firstcrossingsum}).  We will use equations (\ref{EQ:dfcolldS_1storder}) and (\ref{EQ:dfcolldS_2lsl}) to derive the halo bias in the next section.

\section{The linear halo bias}
\label{SEC:halobias}
We now use the results of the preceding sections to obtain the linear halo bias up to leading and next-to-leading orders in (\ref{EQ:sysexpansion}).  The Lagrangian overdensity of halos is

\begin{equation}
\da_h = \frac{ \dd n_c }{ \dd m} \left( \frac{ \dd n}{\dd m}\right)^{-1} - 1 = \frac{\pd \fcoll }{\pd S_s} \left( \frac{\pd \avgfcoll}{\pd S_s} \right)^{-1} -1,
\label{EQ:dh}
\end{equation}
where the ``0" subscript denotes the unconditional first-crossing rate, obtained from evaluating $\pd \fcoll / \pd S_s$ at $S_l =0$ and $\da_l=0$.  Expanding equation (\ref{EQ:dh}) and keeping to linear order in the three-point function yields

\begin{equation}
\begin{split}
\da_h =  \left[ \frac{\pd \fcollgm}{ \pd S_s} \left( \frac{\pd \avgfcollgm}{\pd S_s} \right)^{-1} -1 \right] 
+ \left( \frac{\pd \avgfcollgm}{ \pd S_s} \right)^{-1} \frac{\pd \fcoll^{(1)}}{ \pd S_s} - \left( \frac{\pd \avgfcollgm}{ \pd S_s} \right)^{-2} \frac{\pd \avgfcoll^{( 1)}}{ \pd S_s} \frac{\pd \fcollgm}{ \pd S_s} \\ + \left( \frac{\pd \avgfcollgm}{ \pd S_s} \right)^{-1} \frac{\pd \fcoll^{(2)}}{ \pd S_s} - \left( \frac{\pd \avgfcollgm}{ \pd S_s} \right)^{-2} \frac{\pd \avgfcoll^{(2)}}{ \pd S_s} \frac{\pd \fcollgm}{ \pd S_s}.
\label{EQ:dhexpansion}
\end{split}
\end{equation}
The scale-independent contribution comes from terms containing $\langle \da_s^3 \rangle_c$, while a scale-dependent contribution originates from $\langle \da_s^2 \da_l \rangle_c$ \citep[i.e.][]{2011PhRvD..84f3512D}.  We now treat each case separately.

\subsection{Scale-independent contribution}

First consider the terms up to leading-order, which are given by the first bracketed term and those denoted with superscript of $(1)$ on the right-hand-side of (\ref{EQ:dhexpansion}).  The first bracketed term yields the old excursion set expression for the linear bias with Gaussian initial conditions.  This may be seen by expanding it to first order in $\da_l$, and taking the limit as $S_l\rightarrow 0$,

\begin{equation}
\lim_{S_l\rightarrow0}  \left[ \left( \frac{\pd \avgfcollgm}{\pd S_s} \right)^{-1} \left. \left(\frac{\pd }{\pd \da_l}   \frac{\pd \fcollgm}{ \pd S_s}\right) \right|_{\da_l=0} ~\da_l \right] = \left( \frac{\da_c}{S_s} - \frac{1}{\da_c} \right)~\da_l.
\end{equation}
 The second term in (\ref{EQ:dhexpansion}) gives both scale-dependent and -independent contributions, since it contains both $\langle \da_s^2 \da_l \rangle_c$ and $\langle \da_s^3\rangle_c$, while the third term gives only scale-independent contributions.  Extracting only the terms yielding scale-independent contributions, we have

\begin{equation}
\delta^{(i,1)}_h = \frac{\pd \fcollgm}{\pd S_s}\left(\frac{\pd \avgfcollgm}{\pd S_s} \right)^{-1}\left [ \frac{\pd \langle \da_s^3 \rangle_c}{\pd S_s} \left( \chi -\chi_0 \right) + \langle \da_s^3 \rangle_c~\left( \alpha \chi - \alpha_0 \chi_0 \right)  \right],
\label{EQ:SI_bias_terms_1}
\end{equation}
where the superscript $(i,1)$ denotes leading-order scale-independent terms, and ``0" subscripts on the $\alpha$ and $\chi$ again indicate evaluation at $S_l=0$ and $\da_l=0$.  Following the same procedure used above (see Appendix \ref{APP:SI_bias_1} for details), we obtain the leading-order non-Gaussian scale-independent bias,  

\begin{equation}
\Delta b_i^{(1)}  = -\frac{\mathcal{S}^{(3)}_s}{6} S_s \left[ 2 \frac{\pd \ln \mathcal{S}^{(3)}}{\pd \ln S_s }\left( \frac{1}{\da_c^2} + \frac{1}{S_s}\right)+\frac{1}{\da_c^2} + \frac{3 \da_c^2}{S_s^2} - \frac{2}{S_s} \right],
\label{EQ:SI_bias_1}
\end{equation}
where $\mathcal{S}^{(3)}_s$ is the skewness, $\mathcal{S}^{(3)}_s \equiv \langle \da_s^3\rangle_c / S_s^2$.  Note that this result is identical to equation (115) of \citet{2011PhRvD..84f3512D} for the $N=3$ case.  We have recovered their result from the leading-order excursion set method, where the connected correlators appear in integrals over scales down to the halo scale, but have to a first approximation been treated as constant functions of the smoothing scale.  

Let us now consider the next-to-leading order corrections to (\ref{EQ:SI_bias_1}).  Referring back to equations (\ref{EQ:dfcolldS_2lsl}) and (\ref{EQ:dhexpansion}), the scale-independent contributions come from sub-terms within the last two terms of (\ref{EQ:dhexpansion}) involving either $G^{(1,0,0)}_{s,s,s}$ or $\pd G^{(1,0,0)}_{s,s,s} / \pd S_s$,

\begin{equation}
\delta_h^{(i,2)} = \frac{\pd \fcollgm}{\pd S_s}\left(\frac{\pd \avgfcollgm}{\pd S_s} \right)^{-1} \left[ \frac{3 G^{(1,0,0)}_{s,s,s}}{2} \left( \chi_0 - \chi \right) +  \frac{ \pd G^{(1,0,0)}_{s,s,s}}{\pd S_s} \left( \mu_0 - \mu \right) \right].
\label{EQ:SI_dh_2}
\end{equation}
Following the usual procedure (see Appendix \ref{APP:SI_bias_2}) , we obtain

\begin{equation}
\Delta b_i^{(2)} = \frac{\mathcal{S}^{(3)}_s}{6} S_s \left[  \frac{\pd \ln \mathcal{S}^{(3)}_s}{ \pd \ln S_s} \left( -\frac{7}{ \da_c^2} + \frac{1}{ S_s} \right) -\frac{2}{\da_c^2}+ \frac{2}{S_s} - \frac{2 S_s^2}{ \da_c^2 \mathcal{S}^{(3)}_s} \frac{\pd^2 \mathcal{S}^{(3)}_s}{\pd S_s^2} \right].
\label{EQ:SI_bias_2}
\end{equation}
Note that in addition to corrections to the existing leading-order terms, there is a new contribution involving a second derivative of the skewness.  Finally, adding this result to (\ref{EQ:SI_bias_1}) gives

\begin{equation}
\Delta b_i  = - \frac{\mathcal{S}^{(3)}_s}{6} S_s \left[  \frac{\pd \ln \mathcal{S}^{(3)}_s}{ \pd \ln S_s} \left( \frac{9}{\da_c^2} + \frac{1}{ S_s} \right) +\frac{3}{\da_c^2} + \frac{3 \da_c^2}{S_s^2} -\frac{4}{S_s}+ \frac{2 S_s^2}{ \da_c^2 \mathcal{S}^{(3)}_s} \frac{\pd^2 \mathcal{S}^{(3)}_s}{\pd S_s^2}\right]
\label{EQ:si_bias_2ndorder}
\end{equation}
for the scale-independent bias up to next-to-leading order.

\begin{figure}
\begin{center}
\resizebox{8.8cm}{!}{\includegraphics{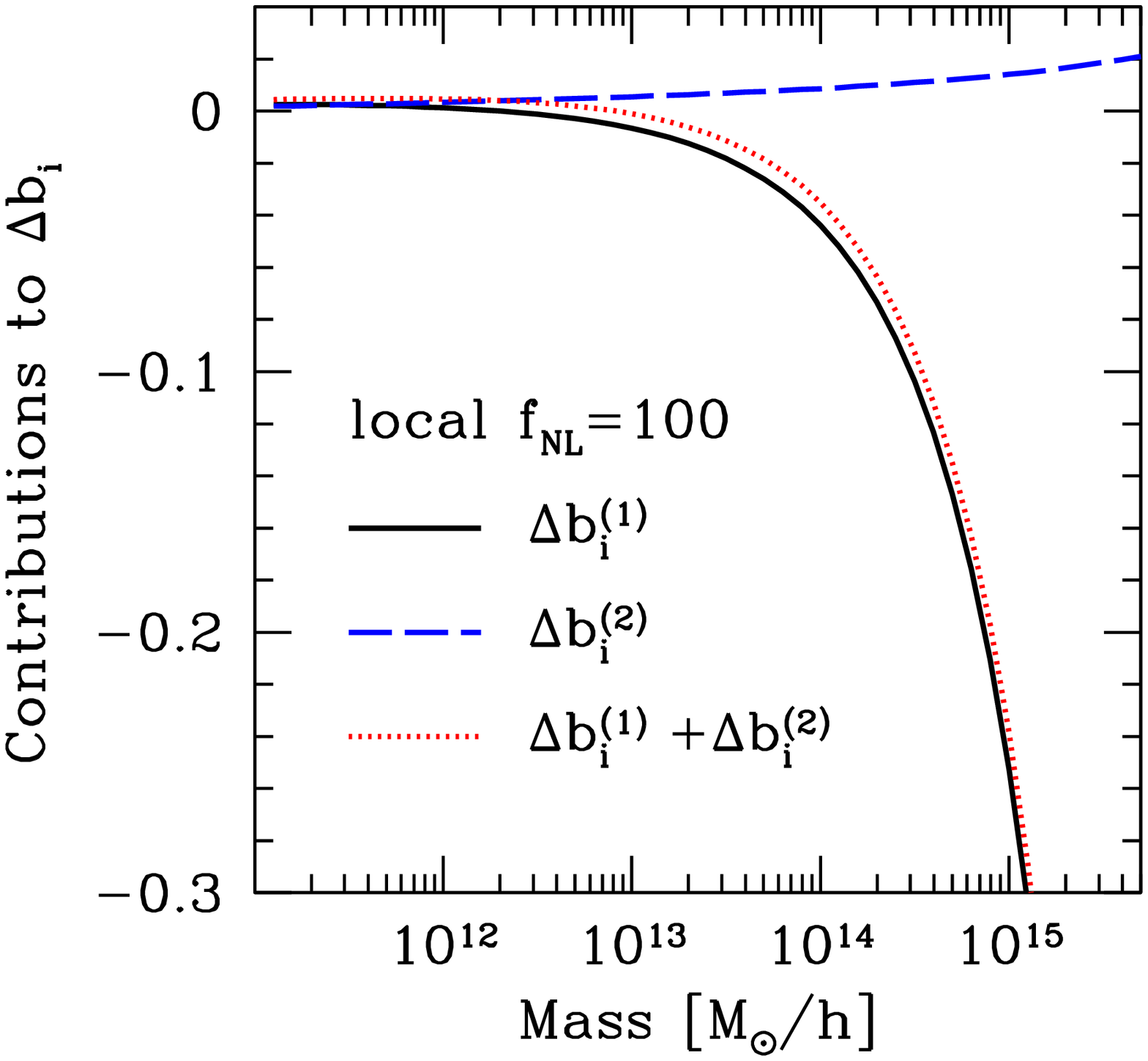}}
\resizebox{8.8cm}{!}{\includegraphics{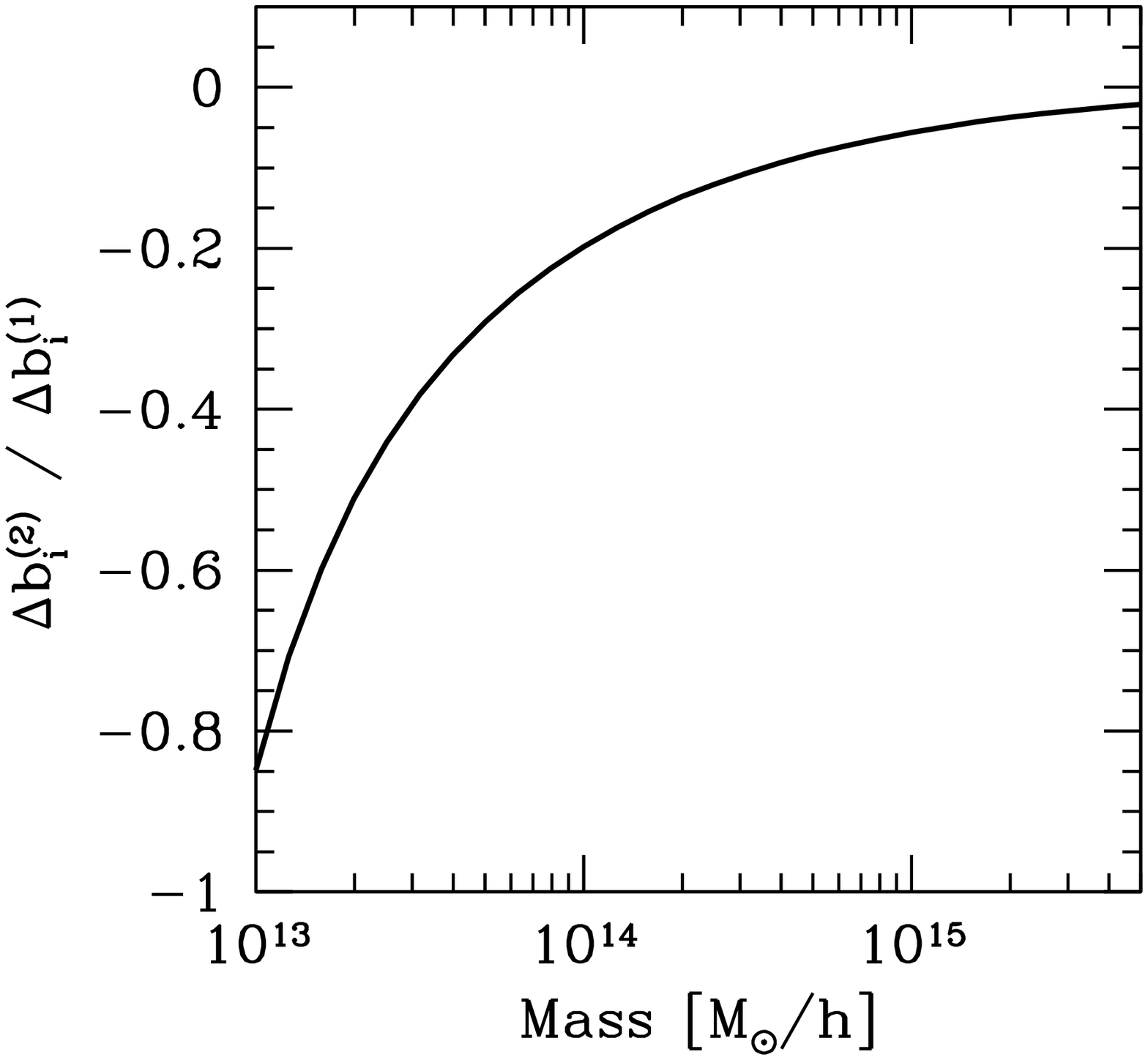}}
\end{center}
\caption{Left panel:  contributions to the scale-independent bias correction at $z=0$ in the local template of PNG as a function of halo mass.  The solid, dashed, and dotted lines show the leading-order, next-to-leading order, and sum respectively.  Right panel:  ratio of the next-to-leading to leading order contributions to the scale-independent bias correction.  }
\label{FIG:SI_bias}
 \end{figure}

Figure \ref{FIG:SI_bias} shows leading and next-to-leading order contributions to the scale-independent bias at $z=0$, as well as their sum for the local template.  Clearly, the next-to-leading order corrections are sub-dominant, relative to leading-order in this case of larger halo masses, suppressing the amplitude by $\sim5\%$ at $M\sim10^{15}\Msun/h$.  The corrections appear to become relatively more important as the mass is decreased.  As we discuss below, this likely signals a breakdown of the expansion, (\ref{EQ:sysexpansion}).

\subsection{Scale-dependent contribution}

The scale-dependent contributions to the bias come from the second and fourth terms in (\ref{EQ:dhexpansion}).  However, equation (\ref{EQ:dhexpansion}) is written in coordinate space, whereas the scale-dependence is manifested in Fourier space.  We employ a convenient trick used by \citet{2011PhRvD..84f3512D} for the conversion.  Their strategy is to take the cross-correlation, $\langle \da_h \da_l \rangle$, between the halo number over-density and the large-scale smoothed density contrast, and rearrange the equations to pick off the bias in Fourier space, $b(k)$.  Although the excursion set model does not formally incorporate a peak constraint, we employ a linear halo bias relation that follows from the correlations between density peaks, $\delta_h(\boldsymbol{k}) = b(k) \delta_s(\boldsymbol{k})$ \citep[see][]{,2010PhRvD..82j3529D,2011PhRvD..84f3512D}.  They showed in the case with Gaussian initial conditions that the peak constraint yields scale-dependent bias corrections proportional to $k^2$ and higher powers of $k$.  We can therefore neglect the effects of the peak constraint since we are focused on the small-$k$ regime.   Using the above bias relation, we write the left-hand side of (\ref{EQ:dhexpansion}) as

\begin{equation}
\langle \da_h \da_l \rangle = \int \frac{\dd^3 \boldsymbol{k}}{(2 \pi)^3} b(k) \MM_s(k) \MM_l(k) P_{\phi}(k).
\label{EQ:lhs_sdbias}
\end{equation}  
If we can now rewrite the appropriate terms on the right-hand side of (\ref{EQ:dhexpansion}) in a similar way, then we can simply read off the bias coefficient in Fourier space. 

Referring again to equations (\ref{EQ:dfcolldS_1storder}) and (\ref{EQ:dhexpansion}), the leading-order scale-dependent contributions come from
\begin{equation}
\da_h^{(d,1)} = \frac{\pd \fcollgm}{\pd S_s}\left(\frac{\pd \avgfcollgm}{\pd S_s} \right)^{-1} \left[ \frac{\pd \langle \da_s^2 \da_l \rangle_c}{\da S_s} \psi  + \langle \da_s^2 \da_l \rangle_c ~\beta \psi - 3 \frac{\pd \langle \da_s^2 \da_l \rangle_c}{\pd S_s}~\chi
-  3 \langle \da_s^2 \da_l \rangle_c~\alpha \chi \right].
\label{EQ:dh_d1}
\end{equation}
The remaining steps are detailed for Appendix (\ref{APP:SD_bias_1}).  In summary, we expand each of the terms to first order in $\da_l$ and take the cross correlation with $\da_l$.  In doing so, the zeroth order terms in each expansion vanish since $\langle \da_l \rangle =0$.  The cross-correlations with the linear terms yield factors of $S_l$.  We then take the limit as $S_l \rightarrow 0$ of the coefficients of the correlators and find that there is a contribution from the first two terms on the right-hand side of (\ref{EQ:dh_d1}), while the last two terms give zero.  Adding up the contributions from the first two terms, we find that

\begin{equation}
 \int \frac{\dd^3 \boldsymbol{k}}{(2 \pi)^3} \Delta b^{(1)}_{d}(k) \MM_s(k) \MM_l(k) P_{\phi}(k) =  \frac{\langle \da_s^2 \da_l \rangle_c}{2 S_s} \left( \frac{\da_c^2}{S_s} - 1\right) + S_s \frac{\pd }{ \pd S_s}\left( \frac{\langle \da_s^2 \da_l\rangle_c}{S_s} \right),
 \label{EQ:sum_d1d2}
\end{equation}
where have expressed the cross-correlation of $\da_l$ with the left-hand side of (\ref{EQ:dh_d1}) using equation (\ref{EQ:lhs_sdbias}).  We can now substitute $\langle \da_s^2 \da_l \rangle_c$ from equation (\ref{EQ:mixedcorrelator}) to obtain

\begin{equation}
\int \frac{\dd^3 \boldsymbol{k}}{(2 \pi)^3} \Delta b^{(1)}_{d}(k) \MM_s(k) \MM_l(k) P_{\phi}(k) = \int \frac{\dd^3 \boldsymbol{k}}{(2 \pi)^3} \MM_s(k) \MM_l(k) P_{\phi}(k) \left[ 2 \delta_c \left( \frac{\da_c}{S_s} - \frac{1}{\delta_c}\right) + 4  \frac{\pd \ln \mathcal{F}_s^{(3)}(k) }{ \pd \ln S_s}  \right] \frac{\mathcal{F}_s^{(3)}(k)}{\MM_s(k)},
\label{EQ:biasint}
\end{equation}
which implies

\begin{equation}
\Delta b^{(1)}_{d}(k) = 2 \left[ \delta_c \left( \frac{\da_c}{S_s} - \frac{1}{\delta_c}\right) + 2 \frac{\pd \ln \mathcal{F}_s^{(3)}(k) }{ \pd \ln S_s} \right] \frac{\mathcal{F}_s^{(3)}(k)}{\MM_s(k)}
\label{EQ:sd_bias_1storder}
\end{equation}
for equation (\ref{EQ:biasint}) to hold for all $k$.  This expression is identical to equation (118) of \citet{2011PhRvD..84f3512D} for the $N=3$ case.  In the local template, the form factor ($\mathcal{F}^{(3)}_s$) approaches $\fNL$ for small $k$, and the first term within the brackets yields the well-known $1/k^2$ form for the scale-dependent bias \citep{2008PhRvD..77l3514D,2008ApJ...677L..77M,2008PhRvD..78l3507A}.  We have also recovered the additional term involving $ \pd \ln \mathcal{F}_s^{(3)}(k) / \pd \ln S_s$, which was only recently pointed out by  \citet{2011PhRvD..84f3512D}.  Although this term is small in the low-$k$ limit in the local template, and therefore of minimal consequence in that case, it is important for other templates as shown in \citet{2011PhRvD..84f1301D}.      

\begin{figure}
\begin{center}
\resizebox{8.5cm}{!}{\includegraphics{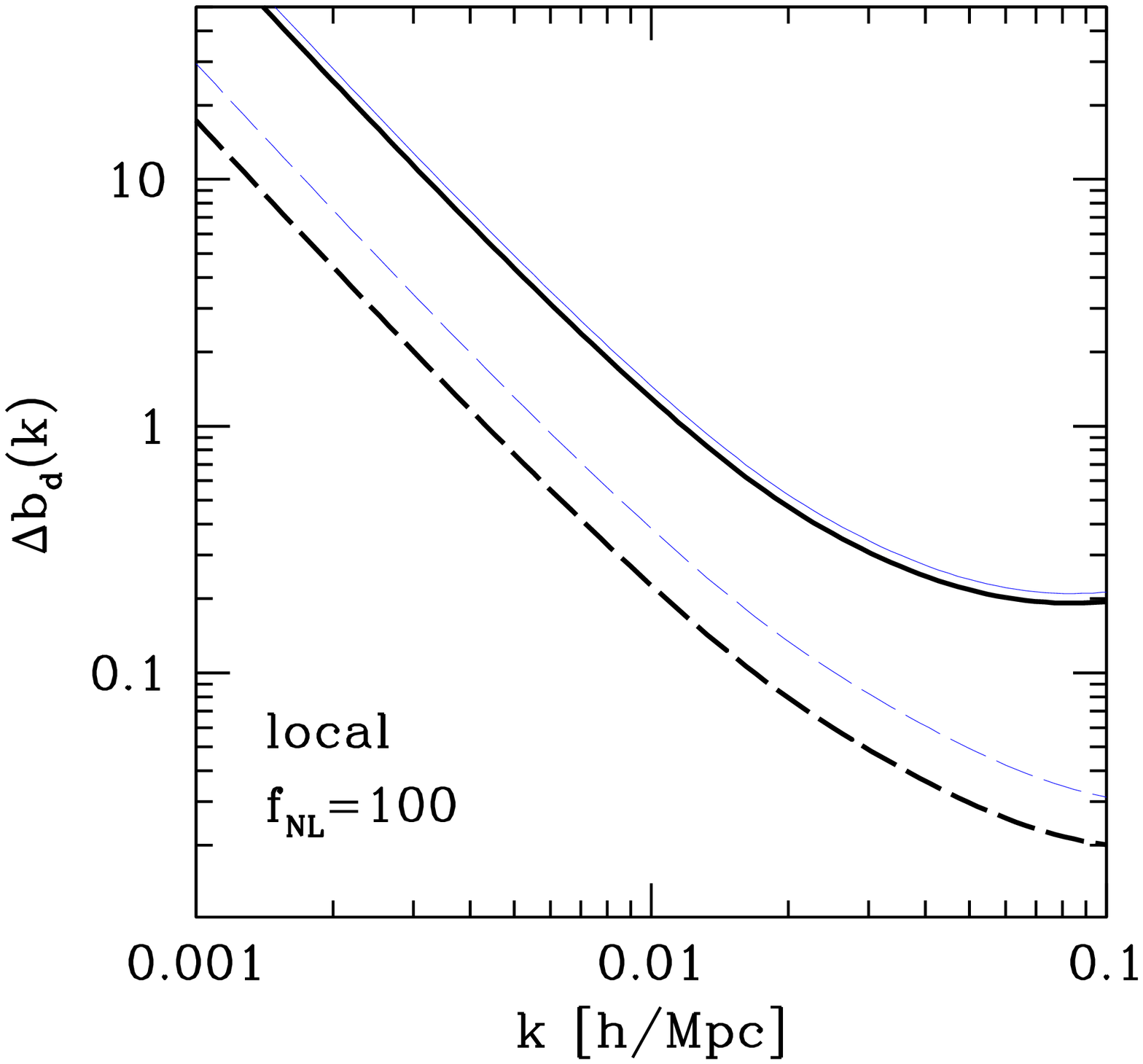}} \hspace{0.13cm}
\resizebox{8.9cm}{!}{\includegraphics{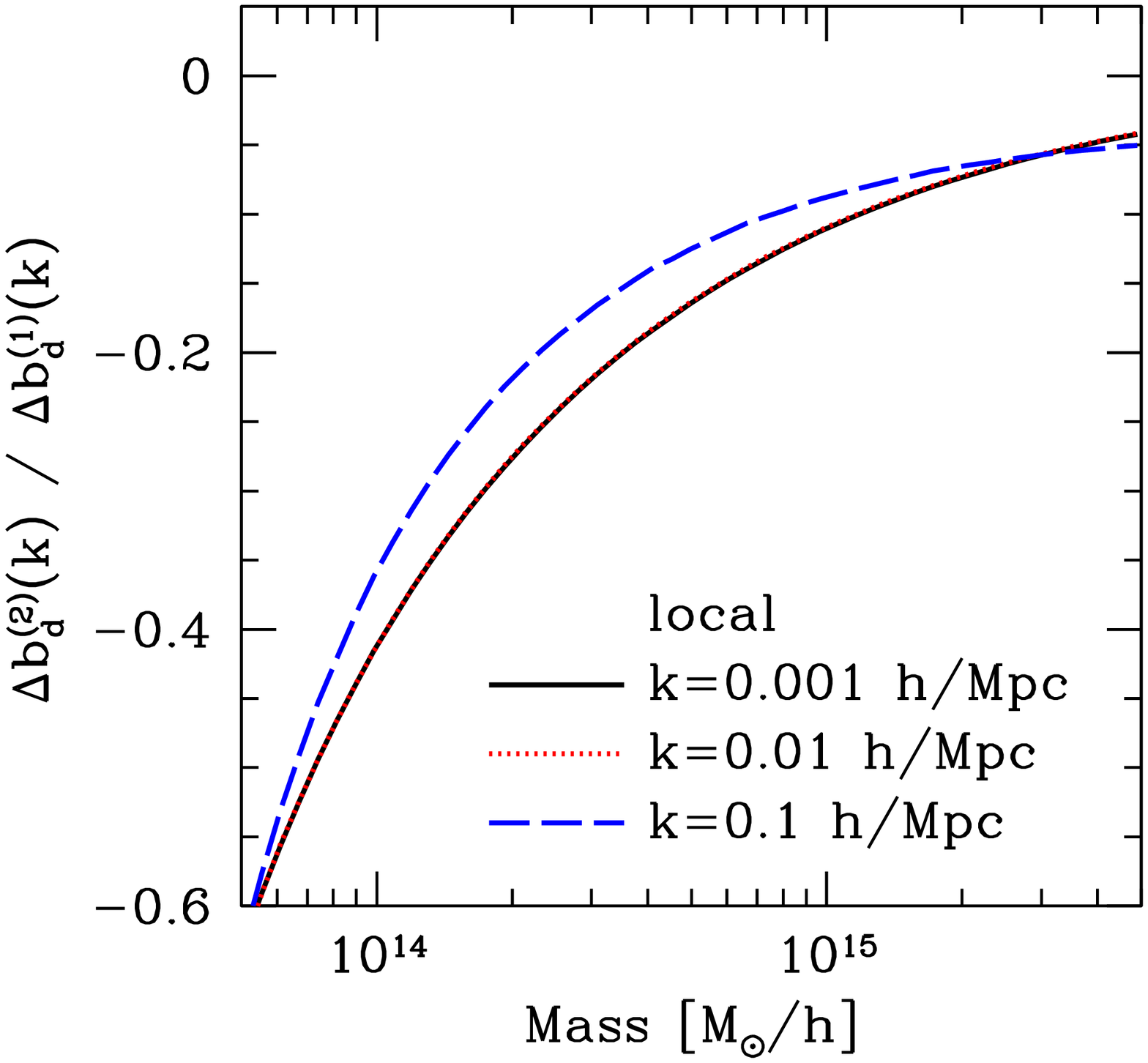}}
\end{center}
\caption{Effects of the next-to-leading order corrections to the scale-dependent halo bias at $z=0$ in the local template of PNG.  Left panel:  the scale-dependent bias to leading order (light and thin) and next-to-leading order (dark and thick) for fixed halo masses of $10^{15} \Msun/h$ (solid) and $10^{14} \Msun/h$ (dashed).  Right panel:  fractional next-to-leading order correction as a function of halo mass for several different wavenumbers.   }
\label{FIG:sd_bias_local}
 \end{figure}

From equations (\ref{EQ:dfcolldS_2lsl}) and (\ref{EQ:dhexpansion}), the next-to-leading order corrections come from

\begin{equation}
\delta_h^{(d,2)} =  \left( \frac{\pd \avgfcollgm}{\pd S_s} \right)^{-1}  \frac{\pd \fcollgm}{\pd S_s} \left[2 \frac{\pd G^{(0,1,0)}_{l,s,s}}{\pd S_s} \mu + 3 G^{(0,1,0)}_{l,s,s} \chi  \right] -  \left( \frac{\pd \avgfcollgm}{\pd S_s} \right)^{-1} \left[ G^{(0,1,0)}_{l,s,s} \frac{\pd \fcollgm}{\pd S_s}  \frac{\da_l }{S_l} + \frac{\pd G^{(0,1,0)}_{l,s,s}}{\pd S_s} \fcollgm \frac{\da_l}{S_l}  \right].
\label{EQ:dh_d2}
\end{equation}
Applying the same procedure as above, the terms in the first set of brackets vanish, while the last two terms give (see Appendix \ref{APP:SD_bias_2})

\begin{equation}
\Delta b^{(2)}_{d}(k) = 2 \left\{ \da_c \left( \frac{-1}{\da_c} \right) - \left[ 1 + \frac{2}{S_s} \left(\frac{\pd \avgfcollgm}{\pd S_s} \right)^{-1} \avgfcollgm \right] \frac{\pd \ln \mathcal{F}^{(3)}_s(k)}{\pd \ln S_s} - S_s \left( \frac{\pd \avgfcollgm }{\pd  S_s}  \right)^{-1} \frac{\avgfcollgm}{\mathcal{F}^{(3)}_s(k)}  \frac{\pd^2 \mathcal{F}^{(3)}_s(k)}{\pd S^2_s} \right\} \frac{\mathcal{F}^{(3)}_s(k)}{\mathcal{M}_s(k)}.
\label{EQ:sd_bias_2}
\end{equation}
We add this result to equation (\ref{EQ:sd_bias_1storder}) to obtain perhaps the main result of this paper; the scale-dependent bias up to next-to-leading order in (\ref{EQ:sysexpansion}),

\begin{equation}
\Delta b_{d}(k) = 2 \left\{ \da_c \left( \frac{\da_c}{S_s} - \frac{2}{\da_c} \right) + \left[ 1 - \frac{2}{ S_s} \left(\frac{\pd \avgfcollgm}{\pd S_s} \right)^{-1} \avgfcollgm \right] \frac{\pd \ln \mathcal{F}^{(3)}_s(k)}{\pd \ln S_s} - S_s \left( \frac{\pd \avgfcollgm }{\pd  S_s}  \right)^{-1} \frac{\avgfcollgm}{\mathcal{F}^{(3)}_s(k)}  \frac{\pd^2 \mathcal{F}^{(3)}_s(k)}{\pd S^2_s} \right\} \frac{\mathcal{F}^{(3)}_s(k)}{\mathcal{M}_s(k)}.
\label{EQ:2ndorderSDbias}
\end{equation}
 
 \begin{figure}
\begin{center}
\resizebox{8.5cm}{!}{\includegraphics{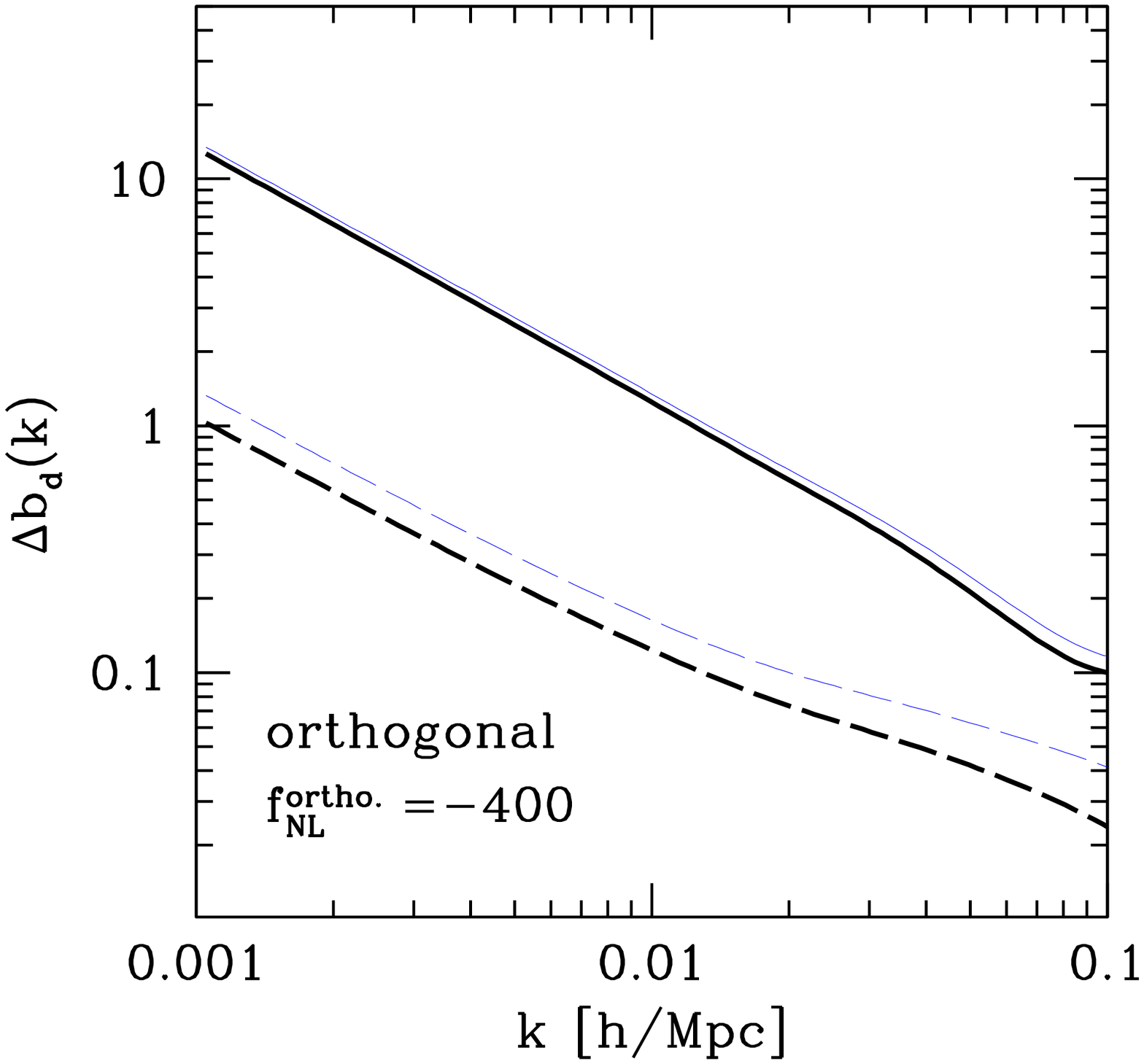}} \hspace{0.13cm}
\resizebox{8.9cm}{!}{\includegraphics{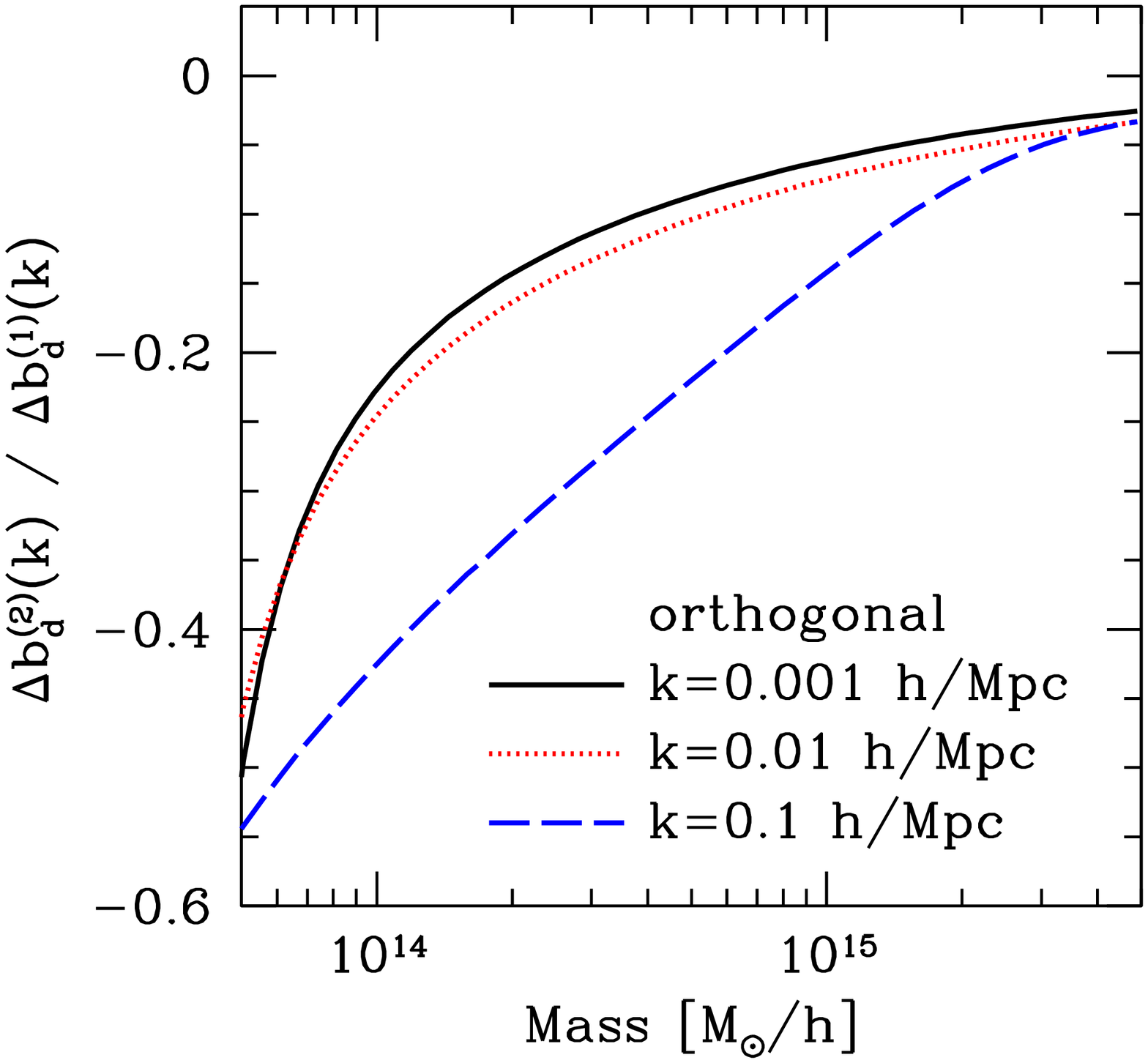}}
\end{center}
\caption{Same as Figure \ref{FIG:sd_bias_local}, but for the orthogonal template of PNG.}
\label{FIG:sd_bias_orthogonal}
 \end{figure}
Figure \ref{FIG:sd_bias_local} shows the scale-dependent bias in the local template to leading order (light and thin) and next-to-leading order (dark and thick) for fixed halo masses of $10^{15} \Msun/h$ (solid) and $10^{14} \Msun/h$ (dashed).  The right panel show the fractional next-to-leading order correction as a function of halo mass for several different wavenumbers.  In the local template, the correction is mainly due to the first term in (\ref{EQ:2ndorderSDbias}).  The other terms play an insignificant role for the wavenumbers considered.  In Figure \ref{FIG:sd_bias_orthogonal}, we reproduce Figure \ref{FIG:sd_bias_local}, but for the orthogonal template, where the other corrections in (\ref{EQ:2ndorderSDbias}) involving derivatives of the form factor make a significant contribution.  The next-to-leading order corrections suppress the amplitude of the scale-dependent bias by $\sim5-10\%$ for halo masses of $M\sim 10^{15}\Msun/h$, and appear to get as large as $\sim20-40\%$ at $M\sim10^{14}\Msun/h$. 

Interestingly, the first term in (\ref{EQ:2ndorderSDbias}) contains a correction that can make a significant difference, even in the low-$k$ limit of the local template, particularly as the halo-mass is decreased.  In this case, the above expression reduces to

\begin{equation}
\Delta b_{d}(k)  \approx \frac{2 \fNL \da_c }{\mathcal{M}_s(k)}\left(\frac{\da_c}{S_s} - \frac{2}{\da_c} \right) = \frac{2 \fNL \da_c }{\mathcal{M}_s(k)}\left( b^E_G - 1 - \frac{1}{\da_c} \right).
\label{EQ:local_lowk_2ndorder}
\end{equation}
When the mass (and therefore $b^{E}_G$) is decreased, the next-to-leading order correction, $1/\da_c$, leads to an increasingly suppressed amplitude compared to the case without correction, as illustrated Figures \ref{FIG:sd_bias_local}. That a supposedly sub-leading correction could make such a big difference in the low-mass regime is likely a sign that the evaluation of the path-integrals through the expansion in equation (\ref{EQ:sysexpansion}) may not be accurate in this regime \citep[see][]{2010ApJ...717..526M}.  We therefore suggest that the scope of equation (\ref{EQ:2ndorderSDbias}) should not be extended to smaller masses; rather, it should be limited to rare halos.  Despite this possible limitation, equation (\ref{EQ:2ndorderSDbias}) illustrates, for the first time, next-to-leading order effects on the scale-dependent bias from PNG within the excursion set formalism.  Finally, we note that the correction in equation (\ref{EQ:local_lowk_2ndorder}) is equal to the merger bias term analytically derived by \cite{2008JCAP...08..031S}.  However, we stress that our correction has nothing to do with mergers, but instead comes from environmental effects encoded in the non-Markovianity from PNG in the excursion set random walks.

\section{Summary and Discussion}
\label{SEC:conclusion}

We have used the path-integral formulation of the excursion set method with constant barrier and sharp $k$-space filter to calculate the conditional collapsed fraction, conditional mass function, and linear halo bias, including both scale-dependent and -independent contributions, in the case of non-Gaussian initial conditions with general bispectra.  Our main results are summarized as follows:

\begin{itemize}

\item{ Equations (\ref{EQ:fcollform}) and (\ref{EQ:fcoll}) give the collapsed fraction to leading order in the expansion of the three-point connected correlator in equation (\ref{EQ:sysexpansion}).   In addition to being useful for other applications, this expression serves as a starting point for deriving the leading-order conditional mass function and  linear halo bias.} 

\item{The next-to-leading order term of the collapsed fraction is given by equation (\ref{EQ:f_coll_2}).  When combined with the leading-order contribution, this term more accurately accounts for the hierarchy of three-point correlators appearing in the excursion-set random-walk distribution function, equation (\ref{EQ:PIgen}). }

\item{Our expression for the leading-order conditional mass function is equation (\ref{EQ:nm_cond}).  We note that this equation recovers the \citet{Lo-Verde:2008rt} mass function in the limit where the environmental density filtering radius is arbitrarily large.  Next-to-leading order corrections in the large-scale environment limit of the conditional mass function can be obtained from equations (\ref{EQ:dfcolldS_1storder}) and (\ref{EQ:dfcolldS_2lsl}). }

\item{  The leading-order scale-independent and -dependent linear bias parameters are given by equation (\ref{EQ:SI_bias_1}) and (\ref{EQ:sd_bias_1storder}) respectively.  The path-integral excursion set method to leading order recovers equations (115) and (118) of \citet{2011PhRvD..84f3512D} for $N=3$.  Notably, we reproduce the additional scale-dependent term involving a derivative of the form factor, which was only recently pointed out by \citet{2011PhRvD..84f3512D}.  While this term is negligible for the local template in the large-scale regime, it is important in other templates, where neglecting it significantly degrades the accuracy when compared to the bias measured from N-body simulations \citep{2011PhRvD..84f1301D}.}

\item{The scale-independent and -dependent bias up to next-to-leading order in (\ref{EQ:sysexpansion}) are given by equations (\ref{EQ:si_bias_2ndorder}) and (\ref{EQ:2ndorderSDbias}) respectively.  For cluster-halo masses ($M\sim10^{15}\Msun/h)$, the next-to-leading order terms suppress the amplitude of both contributions to the bias by $\sim5-10\%$.  In both cases, the relative effects of the next-to-leading order corrections grow as the mass is decreased.  However, we caution that such large effects in the low-mass regime (see Figures \ref{FIG:sd_bias_local} and \ref{FIG:sd_bias_orthogonal} for example) may signal the breakdown of the expansion in (\ref{EQ:sysexpansion}).  We note that even the amplitude of the scale-dependent bias of the local template in the low-$k$ limit is modified in the excursion set prediction, thus altering a well-known result (see equation (\ref{EQ:local_lowk_2ndorder})).     }

\end{itemize}

For the bispectrum templates tested in this work, the next-to-leading order terms become similar to the leading-order terms for masses as large as $M\sim5\times10^{13}\Msun/h$ (for which $\sqrt{S} / \delta_c = 0.62$).  This calls into question the validity of the expansion (\ref{EQ:sysexpansion}) at those masses and below. On the other hand, galaxy surveys aiming to constrain PNG will require predictions that are applicable for galactic halos, whose masses are typically lower.  Corrections to the bias due to environment and formation history may be important if non-Gaussianity is detected, because they could affect the value of $\fNL$ that is extracted.  Our findings therefore motivate more work towards understanding the range of validity of (\ref{EQ:sysexpansion}), and possible alternatives to extend the excursion set method to galactic mass-scales \citep{2011JCAP...02..001D,2011arXiv1108.5512S,2012MNRAS.420..369M}.  Note that we have quantified the environmental effects only due to PNG, for simplicity neglecting additional effects within the model that arise from non-Markovianity introduced by the filter function.  Extensions of this work should focus on the full effects of the environment on halo formation in the context of the halo bias with non-Gaussian initial conditions.    

In addition to resolving the potential issues above, future effort should be devoted to testing the excursion set model against the bias as measured in N-body simulations, along with the numerous analytical predictions in the literature.  We note that some authors have required a ``fudge factor" applied to the analytical predictions in order to reduce their amplitude by $\sim10-25\%$ and improve agreement with simulations \citep[e.g.][]{2009MNRAS.398..321G,2010PhRvD..81f3530G,2010MNRAS.402..191P,2012JCAP...03..002W}.  Though this fudge factor has sometimes been attributed to non-spherical collapse, \cite{2010ApJ...711..907M,2010ApJ...717..515M,2010ApJ...717..526M} note that a similar factor naturally arises within the excursion set model from the stochasticity of the collapse barrier \citep[see also][]{Robertson:2008ad}. We point out that the next-to-leading order corrections that we derived here also act to diminish the amplitude of the scale-dependent bias (i.e by $\lesssim 40 \%$ for $M > 10^{14}\Msun/h$).  Though it is difficult to draw conclusions at this time, it is possible that the ``memory" effects encapsulated in these terms, in combination with other effects (see below), may play a role in the reconciliation of analytical predictions with the lower amplitudes observed in simulations by some authors.  

In comparing the excursion set bias to N-body simulations, it is important to take into account the full range of physical effects accessible to the model:  1) The effects of environment and formation history encoded in the non-Markovianity from both non-Gaussianity \citep[][and the current work]{2010ApJ...717..526M,2011MNRAS.415.1913D,2011MNRAS.418.2403D} and the filter function \citep{2010ApJ...711..907M,2011MNRAS.411.2644M}.  2) The stochasticity of the collapse barrier to parameterize the complex nature of halo collapse \citep{Robertson:2008ad,2010ApJ...717..515M}, and  3) Non-spherical collapse characterized by a dependence of the collapse barrier on the filtering scale \citep{2002MNRAS.329...61S,Zhang:2006ek,2011MNRAS.412.2587D,2011MNRAS.418.2403D,Adshead2012}.  With regards to an improved treatment of the collapse barrier, we became aware of a calculation by \citet{Adshead2012} during the preparation of this manuscript.  While they consider only leading order terms in the expansion, (\ref{EQ:sysexpansion}), they add corrections to the scale-dependent bias due to a moving barrier.  We note that they also recover the $N=3$ case of equation (118) in \citet{2011PhRvD..84f3512D} by setting the barrier to a constant, in agreement with our findings.  The combination of a higher-order treatment of the connected three-point correlators with their moving barrier analysis would be a natural extension to this work.

\section*{Acknowledgments}
We thank Peter Adshead, Vincent Desjacques, Fabian Schmidt, and Eiichiro Komatsu for useful discussions.  We also thank Peter Adshead,  Eric Baxter, Scott Dodelson, and Adam Lidz for sharing a draft of their manuscript with us, and the anonymous referee for helpful comments.  A.D. and D.J. are grateful for the hospitality of the Aspen Center for Physics (supported in part by U.S. NSF grant 1066293) during the completion of the manuscript.  This work was supported in part by U.S. NSF grants AST-0708176 and AST-1009799, and NASA grants NNX07AH09G and NNX11AE09G.

\bibliographystyle{mn2e}
\bibliography{PNGhalobias}


\appendix
\onecolumn
\section{Details of the collapsed fraction calculation}

Since we are interested in both the leading and next-to-leading order terms, we will keep the calculation general for as long as possible, and then specialize to each case.  We note that the approach taken here is based on the appendix of \citet{2011MNRAS.415.1913D} except for the following modification.  Their calculation involves the collapse barrier at two different times, since they were ultimately interested in how primordial non-Gaussianity alters halo collapse epochs.  Here, we modify the calculation to consider the collapse barrier at just a single epoch.  The steps taken here are otherwise identical.

We begin by separating out the terms for which at least one of $i$, $j$, and $k$ is equal to $l$,

\begin{align}
\label{EQ:A}
-\frac{1}{6}\sum_{i,j,k=1}^{s}{\coco~\pd_i \pd_j \pd_k} = & -\frac{1}{6} \sum_{i,j,k=1}^{l-1}  \langle \da_i \da_j \da_k \rangle_c~{\pd_i \pd_j \pd_k}  -\frac{1}{2}  \sum_{i,j=1}^{l-1}{   \langle \da_i \da_j \da_l \rangle_c~\pd_i \pd_j \pd_l}    -\frac{1}{2}  \sum_{i=1}^{l-1}{  \langle \da_i \da^2_l\rangle_c~\pd_i \pd^2_l}  -\frac{1}{6}  \langle \da_l^3\rangle_c~\pd^3_l \\  
\label{EQ:B}
 & -\frac{1}{2}  \sum_{i,j=1}^{l-1}\sum_{k=l+1}^s\langle \da_i \da_j \da_k  \rangle_c~{\pd_i \pd_j \pd_k} -\frac{1}{2}  \sum_{i=1}^{l-1}\sum_{j,k=l+1}^s \langle \da_i \da_j \da_k \rangle_c ~{\pd_i \pd_j \pd_k}  -   \sum_{i=1}^{l-1}\sum_{j=l+1}^s \langle \da_i \da_j \da_l  \rangle_c~{\pd_i \pd_j \pd_l} \\ 
 \label{EQ:C}
 &-\frac{1}{6}  \sum_{i,j,k=l+1}^{s} \langle \da_i \da_j \da_k \rangle_c~{\pd_i \pd_j \pd_k} -\frac{1}{2}  \sum_{i,j=l+1}^{s}\langle \da_i \da_j \da_l \rangle_c~{\pd_i \pd_j \pd_l}   -\frac{1}{2} \sum_{i=l+1}^{s}\langle \da_i \da^2_l \rangle_c ~{\pd_i \pd^2_l}. 
\end{align}
Consider the four terms in (\ref{EQ:A}).  For brevity, we use the following notation:  $\Wgm_{l,s}\equiv \Wgm(\da_l;\da_{l+1},\ldots,\da_s;S_s)$.  Using the factorization property,

\begin{equation}
\Wgm(\da_0;\da_1,\ldots,\da_s;S_s)=\Wgm(\da_0;\da_1,\ldots,\da_l;S_l) \times
 \Wgm(\da_l;\da_{l+1},\ldots,\da_s;S_s),
\label{EQ:factorization}
\end{equation}
we find

\begin{align}
\mathrm{(\ref{EQ:A})}\cdot \Wgm_{0,s}~~ = &~~ -\frac{1}{6}~\Wgm_{l,s}\sum_{i,j,k=1}^{l}{\coco~\pd_i \pd_j \pd_k \Wgm_{0,l}} -\frac{1}{2} ~\pd_l \left(\Wgm_{l,s} \right)\sum_{i,j=1}^{l-1}{\langle \da_{i} \da_j \da_l \rangle_c~\pd_i \pd_j  \Wgm_{0,l}} \nonumber \\  
	& -~\pd_l \left(\Wgm_{l,s}\right)\sum_{i=1}^{l-1}{\langle \da_{i} \da_l^2 \rangle_c~\pd_i \pd_l\left( \Wgm_{0,l}\right)}  -\frac{1}{2} ~\pd^2_l \left( \Wgm_{l,s} \right)\sum_{i=1}^{l-1}{\langle \da_{i} \da_l^2 \rangle_c~\pd_i \Wgm_{0,l}} \nonumber \\ 
	& -\langle \da_l^3 \rangle_c \left[ \frac{1}{6}~\Wgm_{0,l} \pd^3_l\left( \Wgm_{l,s} \right)) +\frac{1}{2} ~ \pd_l \left[ \pd_l\left( \Wgm_{0,l}\right)\pd_l \left( \Wgm_{l,s}\right) \right] \right],
\end{align}
where we have obtained the first term on the right hand side by combining terms.  Plugging this into equation (\ref{EQ:cond_P}) along with the contributions from (\ref{EQ:B}) and (\ref{EQ:C}), and keeping only terms that are first-order in the connected 3-pt correlators yields

\begin{equation}
 P_{\epsilon}(\da_s,S_s|\da_l,S_l) = \PIgm(\da_c;\da_l;\da_s;S_s-S_l)+P_{\epsilon}^{\mathrm{ng}}(\da_s,S_s|\da_l,S_l),
  \label{EQ:P2}
\end{equation}
where we have defined

\begin{equation}
P_{\epsilon}^{\mathrm{ng}}(\da_s,S_s|\da_l,S_l) \equiv \frac{N_a+N_b+N_c}{\PIgm(\da_c ; \da_0; \da_l; S_l-S_0 )},
\end{equation}
with

\begin{align}
\label{EQ:Nagen}
N_a = & \int_{-\infty}^{\da_c}\dd \da_1\ldots \dd \da_{l-1} \int_{-\infty}^{\da_c} \dd \da_{l+1}\ldots \dd \da_{s-1 }  \Biggl[ -\frac{1}{2} ~\pd_l \left(\Wgm_{l,s} \right)\sum_{i,j=1}^{l-1}{\langle \da_{i} \da_j \da_l \rangle_c~\pd_i \pd_j  \Wgm_{0,l}} \\ \nonumber & -\pd_l \left(\Wgm_{l,s}\right)\sum_{i=1}^{l-1}{\langle \da_{i} \da_l^2 \rangle_c~\pd_i \pd_l\left( \Wgm_{0,l}\right)}  -\frac{1}{2} ~\pd^2_l \left( \Wgm_{l,s} \right)\sum_{i=1}^{l-1}{\langle \da_{i} \da_l^2 \rangle_c~\pd_i \Wgm_{0,l}} - \frac{\langle \da_l^3\rangle_c}{6}~\Wgm_{0,l} \pd^3_l\left( \Wgm_{l,s} \right) \\ \nonumber &  -\frac{\langle \da_l^3 \rangle_c}{2} ~ \pd_l \left[ \pd_l\left( \Wgm_{0,l}\right)\pd_l \left( \Wgm_{l,s}\right) \right] \Biggr] 
\\  \nonumber
\\
\label{EQ:Nbgen}
 N_b =  & \int_{-\infty}^{\da_c}\dd \da_1\ldots \dd \da_{l-1} \int_{-\infty}^{\da_c} \dd \da_{l+1}\ldots \dd \da_{s-1 }  \\ 
 & \Biggl[-\frac{1}{2}  \sum_{i,j=1}^{l-1}  \pd_i \pd_j \Wgm_{0,l} \sum_{k=l+1}^s{\coco~\pd_k\Wgm_{l,s}} -\frac{1}{2}~ \sum_{i=1}^{l-1} \pd_i \Wgm_{0,l }\sum_{j,k=l+1}^s{\coco~\pd_j \pd_k \Wgm_{l,s}} \nonumber \\ 
 &  -  \sum_{i=1}^{l-1}  \sum_{j=l+1}^s{ \langle \da_{i} \da_j \da_l\rangle_c~~\pd_l [ \pd_i(\Wgm_{0,l})  \pd_j( \Wgm_{l,s}}) ] \Biggr] \nonumber \\
 \nonumber \\
\label{EQ:Ncgen}
N_c=  & \int_{-\infty}^{\da_c}\dd \da_1\ldots \dd \da_{l-1} \int_{-\infty}^{\da_c} \dd \da_{l+1}\ldots \dd \da_{s-1 } \Biggl[ -\frac{1}{6} ~\Wgm_{0,l}  \sum_{i,j,k=l+1}^{s}{\coco~\pd_i \pd_j \pd_k \Wgm_{l,s}} \\ \nonumber &  -\frac{1}{2}~\pd_l\left( \Wgm_{0,l}   \sum_{i,j=l+1}^{s}{\langle \da_{i} \da_j \da_l\rangle_c~\pd_i \pd_j \Wgm_{l,s}} \right)   -\frac{1}{2} \sum_{i=l+1}^{s}{\langle \da_{i} \da_l^2 \rangle_c~\pd^2_l[  \Wgm_{0,l}  \pd_i   \Wgm_{l,s}}  ] \Biggr]. \nonumber
\end{align}
We cannot evaluate equations (\ref{EQ:Nagen}) through (\ref{EQ:Ncgen}) in their current form, but instead resort to the expansion of the correlators, equation (\ref{EQ:sysexpansion}).  We now consider the leading-order terms in the expansion.

\subsection{The leading order terms}
\label{APP:leadingorder}

The leading-order contributions to $N_a$, $N_b$, and $N_c$ are obtained by substituting the connected correlators evaluated at the endpoints of the sums in (\ref{EQ:simplifiedsum}),

\begin{align}
\label{EQ:Na}
N_a^{(1)} = & \int_{-\infty}^{\da_c}\dd \da_1\ldots \dd \da_{l-1} \int_{-\infty}^{\da_c} \dd \da_{l+1}\ldots \dd \da_{s-1 }  \Biggl[ -\frac{\langle \da_l^3 \rangle_c}{2} ~\pd_l \left(\Wgm_{l,s} \right)\sum_{i,j=1}^{l-1}{\pd_i \pd_j  \Wgm_{0,l}} -\langle \da_l^3 \rangle_c~\pd_l \left(\Wgm_{l,s}\right)\sum_{i=1}^{l-1}{\pd_i \pd_l\left( \Wgm_{0,l}\right)} \ \\  
	& -\frac{\langle \da_l^3 \rangle_c}{2} ~\pd^2_l \left( \Wgm_{l,s} \right)\sum_{i=1}^{l-1}{\pd_i \Wgm_{0,l}} - \frac{\langle \da_l^3\rangle_c}{6}~\Wgm_{0,l} \pd^3_l\left( \Wgm_{l,s} \right) -\frac{\langle \da_l^3 \rangle_c}{2} ~ \pd_l[ \pd_l\left( \Wgm_{0,l}\right)\pd_l\left( \Wgm_{l,s}\right) ] \Biggr]  \nonumber
\\  \nonumber
\\
\label{EQ:Nb}
 N_b^{(1)} =  & \int_{-\infty}^{\da_c}\dd \da_1\ldots \dd \da_{l-1} \int_{-\infty}^{\da_c} \dd \da_{l+1}\ldots \dd \da_{s-1} \Biggl[-\frac{\langle \da_s \da_l^2 \rangle_c}{2}  \sum_{i,j=1}^{l-1}  \pd_i \pd_j \Wgm_{0,l} \sum_{k=l+1}^s{\pd_k\Wgm_{l,s}} \\ & -\frac{\langle \da_s^2 \da_l \rangle_c}{2}~ \sum_{i=1}^{l-1} \pd_i \Wgm_{0,l}\sum_{j,k=l+1}^s{\pd_j \pd_k \Wgm_{l,s}} - \langle \da_s \da_l^2 \rangle_c \sum_{i=1}^{l-1}  \sum_{j=l+1}^s{  \pd_l[  \pd_i (\Wgm_{0,l})  \pd_j (\Wgm_{l,s} )} ] \Biggr] \nonumber
\\ \nonumber
\\
\label{EQ:Nc}
N_c^{(1)}=  & \int_{-\infty}^{\da_c}\dd \da_1\ldots \dd \da_{l-1} \int_{-\infty}^{\da_c} \dd \da_{l+1}\ldots \dd \da_{s-1} \Biggl[ -\frac{\langle \da_s^3 \rangle_c}{6} ~\Wgm_{0,l}  \sum_{i,j,k=l+1}^{s}{\pd_i \pd_j \pd_k \Wgm_{l,s}}  \\ 
 & -\frac{\langle \da_s^2 \da_l \rangle_c}{2}~\pd_l\left( \Wgm_{0,l}   \sum_{i,j=l+1}^{s}{\pd_i \pd_j \Wgm_{l,s}} \right) -\frac{\langle \da_s \da_l^2 \rangle_c}{2}~ \sum_{i=l+1}^{s}{ \pd^2_l [ \Wgm_{0,l} \pd_i \Wgm_{l,s} ]} \Biggr]. \nonumber
\end{align}
We make use of identities similar to equations (48), (49), and (50) of \citet{2010ApJ...717..526M} \cite[see also][]{2010ApJ...711..907M} to evaluate the above expressions.  Specifically, we use the following identities:

\begin{align}
\label{EQ:idA}
  & \pd_c \PIgm(\da_c ; \da_0; \da_l; S_l-S_0 ) =   \sum_{i=1}^{l-1} \int_{-\infty}^{\da_c}\dd\da_1\ldots\dd\da_{l-1}~\pd_i \Wgm_{0,l}  \\ 
  \label{EQ:idB}
   &  \pd_c^2 \PIgm(\da_c ; \da_0; \da_l; S_l-S_0 ) =  \sum_{i,j=1}^{l-1} \int_{-\infty}^{\da_c}\dd\da_1\ldots\dd\da_{l-1}~\pd_i \pd_j \Wgm_{0,l}   \\ 
   \label{EQ:idC}
  & \pd_c^3 \PIgm(\da_c ; \da_0; \da_l; S_l-S_0 ) =  \sum_{i,j,k=1}^{l-1} \int_{-\infty}^{\da_c}\dd\da_1\ldots\dd\da_{l-1}~\pd_i \pd_j \pd_k \Wgm_{0,l}.
 \end{align} 
 Similarly, we would also like to use:
  \begin{align}
 \label{EQ:idD}
   & \pd_c \PIgm(\da_c ; \da_l; \da_s; S_s-S_l ) =   \sum_{i=l+1}^{s-1} \int_{-\infty}^{\da_c}\dd\da_{l+1}\ldots\dd\da_{s-1}~\pd_i \Wgm_{l,s}  \\ 
  \label{EQ:idE}
   &  \pd_c^2 \PIgm(\da_c ; \da_l; \da_s; S_s-S_l ) =  \sum_{i,j=l+1}^{s-1} \int_{-\infty}^{\da_c}\dd\da_{l+1}\ldots\dd\da_{s-1}~\pd_i \pd_j \Wgm_{l,s}   \\ 
   \label{EQ:idF}
  & \pd_c^3 \PIgm(\da_c ; \da_l; \da_s; S_s-S_l ) =  \sum_{i,j,k=l+1}^{s-1} \int_{-\infty}^{\da_c}\dd\da_{l+1}\ldots\dd\da_{s-1}~\pd_i \pd_j \pd_k \Wgm_{l,s}.
\end{align}
Note, however, that the summations in equations (\ref{EQ:Nb}) and (\ref{EQ:Nc}) are up to $s$ and not to $s-1$.  The latter is required to use equations (\ref{EQ:idD}), (\ref{EQ:idE}), and (\ref{EQ:idF}).  As \citet{2010ApJ...717..526M} point out, this is not a problem since we are ultimately interested in calculating $f_{\mathrm{coll}}(S_s|\da_l,S_l)$, which is given by 

\begin{equation}
f_{\mathrm{coll}}(S_s|\da_l,S_l) = 1 - \int_{-\infty}^{\da_c}{\dd\da_s~ \PIgm(\da_c;\da_l;\da_s;S_s-S_l)} -  \int_{-\infty}^{\da_c}{\dd\da_s~ P_{\epsilon}^{\mathrm{ng}}(\da_s,S_s|\da_l,S_l)}.
 \end{equation}
 We will therefore evaluate $\int_{-\infty}^{\da_c}\dd \da_s~N_{a(bc)}$ instead of $N_{a(bc)}$. In that case, we can use
 
\begin{align}
 \label{EQ:idG}
   & \pd_c U_{\epsilon}(l,s) =   \sum_{i=l+1}^{s} \int_{-\infty}^{\da_c}\dd\da_{l+1}\ldots\dd\da_{s}~\pd_i \Wgm_{l,s}  \\ 
  \label{EQ:idH}
   &  \pd_c^2 U_{\epsilon}(l,s)  =  \sum_{i,j=l+1}^{s} \int_{-\infty}^{\da_c}\dd\da_{l+1}\ldots\dd\da_{s}~\pd_i \pd_j \Wgm_{l,s}   \\ 
   \label{EQ:idI}
  & \pd_c^3 U_{\epsilon}(l,s)  =  \sum_{i,j,k=l+1}^{s} \int_{-\infty}^{\da_c}\dd\da_{l+1}\ldots\dd\da_{s}~\pd_i \pd_j \pd_k \Wgm_{l,s},
  \end{align}
where we have defined 
  
  \begin{align}
 U_{\epsilon}(l,s) = \int_{-\infty}^{\da_c}\dd \da_s \PIgm(\da_c ; \da_l;\da_s ; S_s-S_l )
  \end{align}
 
 The strategy is to substitute the right-hand sides of equations (\ref{EQ:idA}) - (\ref{EQ:idC}) and (\ref{EQ:idG}) - (\ref{EQ:idI}) wherever they appear in equations (\ref{EQ:Na}), (\ref{EQ:Nb}), and (\ref{EQ:Nc}).  Equation (\ref{EQ:Na}) produces
 
\begin{align}
\int_{-\infty}^{\da_c}\dd \da_s~N^{(1)}_{a} =  & -\frac{ \langle \da^3_l \rangle_c}{2}~\pd_c^2\left[ \PIgm(0,l)\right] \pd_l\left[ U_{\epsilon}(l,s)\right] 
-\langle \da_l^3 \rangle_c~ \pd_l\pd_c \left[  \PIgm(0,l) \right] \pd_l \left[ U_{\epsilon}(l,s)\right] 
-\frac{ \langle \da^3_l \rangle_c}{2}~\pd_c\left[ \PIgm(0,l)\right] \pd_l^2 \left[ U_{\epsilon}(l,s)\right] \nonumber \\
& -\frac{ \langle \da^3_l \rangle_c}{6}~ \PIgm(0,l) \pd^3_l\left[ U_{\epsilon}(l,s)\right]
  -\frac{ \langle \da^3_l \rangle_c}{2}~\pd_l^2\left[ \PIgm(0,l)\right] \pd_l\left[ U_{\epsilon}(l,s)\right]  
-\frac{ \langle \da^3_l \rangle_c}{2}~ \pd_l \PIgm(0,l)  \pd^2_l\left[ U_{\epsilon}(l,s)\right].
\label{EQ:intNa}
\end{align}
For brevity, we have used the shorthand notation:  $\PIgm(0,l)\equiv\PIgm(\da_c,\da_l-\da_0,S_l-S_0)$.  Following a similar procedure for $N_b$ and $N_c$ yields

\begin{align}
\int_{-\infty}^{\da_c}\dd \da_s~N^{(1)}_{b} =  -\frac{ \langle \da_s \da^2_l  \rangle_c}{2}~\pd_c^2\left[ \PIgm(0,l)\right] \pd_c \left[ U_{\epsilon}(l,s)\right]
 -\frac{ \langle \da_l \da_s^2 \rangle_c}{2}~\pd_c\left[ \PIgm(0,l)\right] \pd_c^2 \left[ U_{\epsilon}(l,s)\right] & - \langle \da_s \da^2_l \rangle_c~\pd_l \pd_c \left[ \PIgm(0,l) \right]  \pd_c\left[ U_{\epsilon}(l,s)\right] \nonumber \\
  - & \langle \da_s \da^2_l \rangle_c~\pd_c  \left[ \PIgm(0,l) \right]  \pd_l \pd_c\left[ U_{\epsilon}(l,s)\right]
\label{EQ:intNb}
 \end{align}
 and
 \begin{align}
\int_{-\infty}^{\da_c}\dd \da_s~N^{(1)}_{c} =  & -\frac{ \langle \da^3_s \rangle_c}{6}~ \PIgm(0,l) \pd_c^3\left[ U_{\epsilon}(l,s)\right] 
-\frac{\langle \da_s^2 \da_l \rangle_c}{2}~  \PIgm(0,l) \pd_l \pd_c^2 \left[ U_{\epsilon}(l,s)\right] 
-\frac{ \langle \da^2_s \da_l \rangle_c}{2}~\pd_l\left[ \PIgm(0,l)\right] \pd_c^2 \left[ U_{\epsilon}(l,s)\right] \nonumber \\
& -\frac{ \langle \da_s \da^2_l \rangle_c}{2}~ \PIgm(0,l) \pd_c \pd^2_l\left[ U_{\epsilon}(l,s)\right]
  -\frac{ \langle \da_s \da_l^2 \rangle_c}{2}~\pd_l^2\left[ \PIgm(0,l)\right] \pd_c\left[ U_{\epsilon}(l,s)\right]  
-\langle \da_s \da^2_l \rangle_c~ \pd_l \left[ \PIgm(0,l) \right]  \pd_l\pd_c\left[ U_{\epsilon}(l,s)\right].
\label{EQ:intNc}
\end{align}
The next step is to add equations (\ref{EQ:intNa}), (\ref{EQ:intNb}), and (\ref{EQ:intNc}) and take the continuum limit.   Note that if we send $\epsilon \rightarrow 0$, then
\begin{equation}
U_{\epsilon=0}(l,s) = \erf\left( (\da_c-\da_l)/\sqrt{2(S_s-S_l)} \right).
\label{EQ:Ueq}
\end{equation}
We make use of the following properties:

\begin{align}
& \pd_l^3  U_{\epsilon=0}(l,s) = - \pd_c^3  U_{\epsilon=0}(l,s)  = \pd_l \pd_c^2 U_{\epsilon=0}(l,s) = - \pd^2_l \pd_c U_{\epsilon=0}(l,s)\\ \nonumber \\
& \pd_l^2  U_{\epsilon=0}(l,s) =  \pd_c^2  U_{\epsilon=0}(l,s)  =  - \pd_l \pd_c U_{\epsilon=0}(l,s) \\ \nonumber \\
& \pd_l  U_{\epsilon=0}(l,s)  =  -\pd_c U_{\epsilon=0}(l,s). 
\end{align}
After some rearranging, we find

\begin{align}
\int_{-\infty}^{\da_c}{\dd\da_s~ P_{\epsilon=0}^{\mathrm{ng}}(\da_s,S_s|\da_l,S_l)} = 
-\mathcal{A}(S_l,S_s)~ \frac{\pd^3_c U_{\epsilon=0}(l,s)}{6} 
- \mathcal{B}(S_l,S_s)~\left\{ \frac{\pd_c \left[ \PIgmF(0,l) \right]+\pd_l \left[ \PIgmF(0,l)\right] }{\PIgmF(0,l)} \right\} \frac{\pd^2_c U_{\epsilon=0}(l,s)}{2} \nonumber \\ 
- \mathcal{C}(S_l,S_s)~\left\{ \frac{\pd^2_c \left[ \PIgmF(0,l) \right] + \pd^2_l \left[ \PIgmF(0,l) \right] + 2 \pd_l \pd_c \left[ \PIgmF(0,l)\right]  }{\PIgmF(0,l)} \right\} \frac{\pd_c U_{\epsilon=0}(l,s)}{2},
\label{EQ:2ndtolast}
\end{align}
where $\Afd$, $\Bfd$, and $\Cfd$ are defined in equations (\ref{EQ:Afunc}) through (\ref{EQ:Cfunc}).  Equation (\ref{EQ:2ndtolast}) may be evaluated using (\ref{EQ:Ueq}) and the probability density $\PIgmF(0,l)$, which is given by equation (\ref{EQ:PIGM}).  Finally, performing these substitutions, differentiating, and combining with the Gaussian and Markovian term yields equation (\ref{EQ:fcoll}).

\subsection{The next-to-leading order terms}
\label{APP:nextoleading}

Referring back to equations (\ref{EQ:Nagen}) through (\ref{EQ:Ncgen}), we substitute the appropriate next-to-leading order terms from (\ref{EQ:sysexpansion}) to obtain
\begin{align}
\label{Na2}
 N_a^{(2)} = & \int_{-\infty}^{\da_c}\dd \da_1\ldots \dd \da_{l-1} \int_{-\infty}^{\da_c} \dd \da_{l+1}\ldots \dd \da_{s-1} \left[-\frac{G^{(1,0,0)}_{l,l,l}}{2}\partial_l(\Wls)\sum_{i,j=1}^{l-1}(S_i+S_j-2S_l)\partial_i\partial_j\Wl \right.  \\ & \left. - G^{(1,0,0)}_{l,l,l}\partial_l(\Wls)\sum_{i=1}^{l-1}(S_i-S_l)\partial_i\partial_l\Wl-\frac{G^{(1,0,0)}_{l,l,l}}{2}\partial_l^2(\Wls)\sum_{i=1}^{l-1}(S_i-S_l)\partial_i\Wl\right]
\nonumber \\
\label{Nb2}
N^{(2)}_b = &  \int_{-\infty}^{\da_c}\dd \da_1\ldots \dd \da_{l-1} \int_{-\infty}^{\da_c} \dd \da_{l+1}\ldots \dd \da_{s-1} \nonumber \\ & \left\{-\frac{1}{2}\sum_{i,j=1}^{l-1}\sum_{k=l+1}^{s}\left[(S_i+S_j-2S_l)G^{(1,0,0)}_{l,l,s}+(S_k-S_s) G^{(0,0,1)}_{l,l,s} \right]\partial_i\partial_j(\Wl)\partial_k(\Wls)\right. \\ \nonumber
-&\frac{1}{2}\sum_{i=1}^{l-1}\sum_{j,k=l+1}^{s}\left[(S_i-S_l)G^{(1,0,0)}_{l,s,s}+(S_j+S_k-2S_s) G^{(0,1,0)}_{l,s,s} \right]\partial_i(\Wl)\partial_j\partial_k(\Wls)\\ \nonumber
-&\left.\sum_{i=1}^{l-1}\sum_{j=l+1}^{s}\left[(S_i-S_l)G^{(1,0,0)}_{l,l,s} +(S_j-S_s) G^{(0,0,1)}_{l,l,s}\right]\partial_l\left[\partial_i(\Wl)\partial_j(\Wls)\right]\right\}
\nonumber \\
\label{Nc2}
N_c^{(2)} =&\int_{-\infty}^{\da_c}\dd \da_1\ldots \dd \da_{l-1} \int_{-\infty}^{\da_c} \dd \da_{l+1}\ldots \dd \da_{s-1} \left\{-\frac{G^{(1,0,0)}_{s,s,s}}{6}\Wl\sum_{i,j,k=l+1}^{s}(S_i+S_j+S_k-3S_s)\partial_i\partial_j\partial_k\Wls\right. \\ \nonumber
-&\frac{G^{(0,1,0)}_{l,s,s}}{2}\partial_l\left[\Wl\sum_{i,j=l+1}^{s}(S_i+S_j-2S_s)\partial_i\partial_j\Wls\right] -\left.\frac{G^{(0,0,1)}_{l,l,s}}{2}\partial_l^2\left[\Wl\sum_{i=l+1}^{s}(S_i-S_s)\partial_i\Wls\right]\right\}.
\end{align}
The strategy is similar to our procedure in the last section.  As before, we will for convenience evaluate $\int_{-\infty}^{\da_c} \dd \da_s N_{a (bc)}$ rather than the above expressions.  However, there are some additional ingredients in the calculation due to the the presence of the $S_i$ in the summations.  We will encounter the quantities

\begin{equation}
\label{Gammaa}
\Gamma_a = \intc\ddl\sum_{i=1}^{l-1}S_i\partial_i\Wl,
\end{equation}
and 

\begin{equation}
\label{Gammab}
\Gamma_b = \intc d\delta_s\intc\ddss\sum_{k=l+1}^{s}S_k\partial_k\Wls .
\end{equation}
The former may be easily evaluated by noting that
\begin{align}
\Gamma_a = \sum_{i=1}^{l-1}S_i \int_{-\infty}^{\da_c} \dd \delta_1 \ldots \dd \delta_{l-1 } \pd_i \Wgm =  &  \sum_{i=1}^{l-1}S_i \int_{-\infty}^{\da_c} \dd \delta_1 \ldots \delta_{i-1} \Wgm(\da_0; \da_1,\ldots, \da_{i-1},\da_c; S_i) \nonumber \\ & \times \int_{-\infty}^{\da_c}\dd \delta_{i+1} \ldots \dd \delta_{l-1 } \Wgm(\da_c; \da_{i+1},\ldots, \da_{l}; S_l-S_i) \nonumber \\  = &  \sum_{i=1}^{l-1}S_i \PIgm(0; \da_c; S_i) \PIgm(\da_c; \da_l; S_l-S_i),
\label{EQ:Gamma_a_sum}
\end{align}
where \citep[see equations (79) and (80) of][]{2010ApJ...711..907M}

\begin{equation}
\PIgm(0; \da_c; S_i) = \sqrt{\epsilon} \frac{1}{\sqrt{\pi}} \frac{\da_c}{S_i^{3/2}} \exp\left[ -\frac{\da_c^2}{2 S_i} \right],
\end{equation}
and

\begin{equation}
\PIgm(\da_c; \da_l; S_l-S_i) = \sqrt{\epsilon} \frac{1}{\sqrt{\pi}} \frac{\da_c - \da_l}{(S_l-S_i)^{3/2}} \exp\left[ -\frac{(\da_c-\da_l)^2}{2 (S_l-S_i)} \right].
\end{equation}
Noting that $\PIgm(0; \da_c; S_i) \PIgm(\da_c; \da_l; S_l-S_i)$ in the sum remains finite for $\epsilon \rightarrow 0$, in the continuum limit we may replace the sum with an integral,

\begin{equation}
\sum_{i=1}^{l-1} \epsilon \rightarrow \int_{0}^{S_l} \dd S_i,
\end{equation}
so that

\begin{align}
\Gamma_a =   \frac{\delta_c(\delta_c-\delta_l)}{\pi}\int_{0}^{S_l}dS_i\frac{S_i}{S_i^{3/2}(S_l-S_i)^{3/2}}\exp\left[-\frac{\delta_c^2}{2S_i}-\frac{(\delta_c-\delta_l)^2}{2(S_l-S_i)}\right].
\label{EQ:gam_a_explicit}
\end{align}
Using equation (108) of \citet{2010ApJ...711..907M}, we obtain

\begin{equation}
\Gamma_a = \sqrt{\frac{2}{\pi}}\frac{\delta_c}{\sqrt{S_l}}\exp\left[-\frac{(2\delta_c-\delta_l)^2}{2S_l}\right].
\end{equation}
For $\Gamma_b$, we may write 

\begin{equation}
\Gamma_b =\intc d\delta_s\left[\frac{\partial}{\partial\delta_s}S_s\PIgm(\delta_l;\delta_s;S_s-S_l)+\sum_{k=l+1}^{s-1}S_k\PIgm(\delta_l;\delta_c;S_k-S_l)\PIgm(\delta_c;\delta_s;S_s-S_k)\right].
\end{equation}
The first term evaluates to zero and, after converting the sum to an integral over $\dd S_k$, the second terms gives

\begin{equation}
\Gamma_b=\frac{1}{\pi}\intc d\delta_s\int_{S_l}^{S_s}dS_kS_k\frac{(\delta_c-\delta_l)(\delta_c-\delta_s)}{(S_k-S_l)^{3/2}(S_s-S_k)^{3/2}}\exp\left[-\frac{(\delta_c-\delta_l)^2}{2(S_k-S_l)}-\frac{(\delta_c-\delta_s)^2}{2(S_s-S_k)}\right],
\end{equation}
which can be evaluated to

\begin{equation}
\Gamma_b=(\delta_c-\delta_l)\mathrm{erfc}\left[\frac{\delta_c-\delta_l}{\sqrt{2(S_s-S_l)}}\right]+\sqrt{\frac{2}{\pi}}\frac{S_l}{\sqrt{S_s-S_l}}\exp\left[-\frac{(\delta_c-\delta_l)^2}{2(S_s-S_l)}\right].
\label{EQ:gam_b_explicit}
\end{equation}
Equations (\ref{EQ:gam_a_explicit}) and (\ref{EQ:gam_b_explicit}) will be needed to evaluate the final expression for the next-to-leading order contribution.

Utilizing again the procedure of replacing sums of partial derivatives with partial derivatives of the probability density, and cumulative probability (i.e. equations (\ref{EQ:idA}) through (\ref{EQ:idC}) and (\ref{EQ:idG}) through (\ref{EQ:idI})), we obtain

\begin{align}
\label{dNa_final}
\intc d\delta_s N^{(2)}_a = & -G^{(1,0,0)}_{l,l,l} \left\{ \partial_l\Ue\left[ \pd_c \Gamma_a+ \pd_l \Gamma_a-S_l\partial_c^2\Pe-S_l\partial_c\partial_l\Pe\right] \right. \\  & \left. -\frac{\partial_l^2\Ue}{2}\left[\Gamma_a-S_l\partial_c\Pe\right] \right\}
\nonumber \\
\label{dNb_final}
\intc d\delta_s N^{(2)}_b = &- G^{(1,0,0)}_{l,l,s}\partial_c\Ue\left[ \pd_c \Gamma_a-S_l\partial_c^2\Pe\right]-\frac{G^{(0,0,1)}_{l,l,s}}{2}\partial_c^2\Pe\left[\Gamma_b-S_s\partial_c\Ue\right]  \\
-&\frac{G^{(1,0,0)}_{l,s,s}}{2}\partial_c^2\Ue\left[\Gamma_a-S_l\partial_c\Pe\right]- G^{(0,1,0)}_{l,s,s}\partial_c\Pe\left[ \pd_c \Gamma_b-S_s\partial_c^2\Ue\right]  \nonumber \\ 
-& G^{(1,0,0)}_{l,l,s} \partial_l\left[\left(\Gamma_a-S_l\partial_c\Pe\right)\partial_c\Ue\right]-  G^{(0,0,1)}_{l,l,s}\partial_l\left[\partial_c\Pe\left(\Gamma_b-S_s\partial_c\Ue\right)\right]
\nonumber \\
\label{dNc_final}
\intc d\delta_s N^{(2)}_c =&-\frac{G^{(1,0,0)}_{s,s,s}}{2}\Pe\left( \pd_c^2 \Gamma_b-S_s\partial_c^3\Ue\right)- G^{(0,1,0)}_{l,s,s}\partial_l\left[\Pe\left( \pd_c \Gamma_b-S_s\partial_c^2\Ue\right)\right]   \\ 
-&\frac{G^{(0,0,1)}_{l,l,s}}{2}\partial_l^2\left[\Pe\left(\Gamma_b-S_s\partial_c\Ue\right)\right] \nonumber
\end{align}
Plugging in $\Gamma_a$ and $\Gamma_b$ along with ${\PIgmF}(0,l)$ and $U_{\epsilon=0}(l,s)$ from the last section, summing the contributes from $N_a$, $N_b$ and $N_c$, and manipulating the result yields equation (\ref{EQ:f_coll_2}) - the next-to-leading order contribution to the collapsed fraction.

\section{Details of the linear halo bias calculation}

\subsection{Scale-independent terms}

\subsubsection{Leading-order}
\label{APP:SI_bias_1}

We would like to write our final result in terms of the skewness ($\mathcal{S}^{(3)}_s = \langle \da_s^3\rangle_c/S_s^2$) rather than the three-point function, so we first convert derivatives of $\langle \da_s^3\rangle_c$ to derivatives of $\mathcal{S}^{(3)}_s$ using 

\begin{equation}
\frac{\pd \langle \da_s^3 \rangle_c}{\pd S_s} = S_s^2 \frac{\pd \mathcal{S}^{(3)}_s }{\pd S_s} + \frac{2}{S_s} \langle \da_s^3 \rangle_c.
\end{equation} 
Let us consider the following three terms from equation (\ref{EQ:SI_bias_terms_1}) separately:

\begin{equation}
 I_1= S_s^2 \frac{\pd \mathcal{S}^{(3)}_s}{\pd S_s} \frac{\pd \fcollgm}{\pd S_s}\left(\frac{\pd \avgfcollgm}{\pd S_s} \right)^{-1} \left( \chi  -\chi_0\right)
 \label{EQ:I1}
\end{equation}

\begin{equation}
I_2 = \frac{2}{S_s} \langle \da_s^3 \rangle_c   \frac{\pd \fcollgm}{\pd S_s}\left(\frac{\pd \avgfcollgm}{\pd S_s} \right)^{-1}   \left(  \chi  -\chi_0\right).
 \label{EQ:I2}
 \end{equation}

\begin{equation}
 I_3 = \langle \da_s^3\rangle_c   \frac{\pd \fcollgm}{\pd S_s}\left(\frac{\pd \avgfcollgm}{\pd S_s} \right)^{-1}  \left( \alpha \chi - \alpha_0\chi_0 \right)
 \label{EQ:I3}
 \end{equation}
 Starting with (\ref{EQ:I1}), we expand to first order in $\da_l$ about $\da_l=0$ and take the limit as $S_l\rightarrow0$ to obtain

\begin{equation}
I_1 =  S^2_s \frac{\pd \mathcal{S}^{(3)}_s}{\pd S_s} \lim_{S_l\rightarrow0} \left( \left. \frac{\pd \chi}{\pd \da_l}\right|_{\da_l=0} \right)\da_l = - S_s^2 \frac{\pd \mathcal{S}^{(3)}_s}{\pd S_s}\frac{1}{3} \left( \frac{1}{\da_c^2} + \frac{1}{S_s}\right) \da_l.
 \label{EQ:lima}
\end{equation}
Note that equation (\ref{EQ:I2}) is similar to (\ref{EQ:I1}) so that

\begin{equation}
I_2 = -\frac{2}{S_s}\frac{\langle \da_s^3 \rangle_c}{3} \left( \frac{1}{\da_c^2} + \frac{1}{S_s}\right)~\delta_l.
\end{equation}  
Finally, equation (\ref{EQ:I3}) yields

\begin{equation}
 I_3 = \langle \da_s^3 \rangle_c \lim_{S_l \rightarrow 0} \left( \left. \frac{\pd (\alpha \chi)}{\pd \da_l}\right|_{\da_l=0} \right) \da_l= \frac{\langle \da_s^3 \rangle_c}{2 Ss} \left( -\frac{\da_c^2}{S_s^2} +\frac{2}{S_s} +\frac{1}{\da_c^2} \right)~\da_l. 
\end{equation}
Rewriting in terms of the skewness and combing $I_2$ and $I_3$, we find

\begin{equation}
I_2+I_3 = - \frac{\mathcal{S}^{(3)}}{6}~S_s\left( \frac{1}{\da_c^2} + \frac{3 \da_c^2}{S_s^2} - \frac{2}{S_s} \right) \da_l.
\end{equation}
Combining this result with $I_1$ gives equation (\ref{EQ:SI_bias_1}).

\subsubsection{Next-to-leading order}
\label{APP:SI_bias_2}

Following the same procedure as in the last section, equation (\ref{EQ:SI_dh_2}) to first order in $\da_l$ and, in the limit $S_l \rightarrow 0$, becomes

\begin{equation}
\delta_h^{(2,i)} =  \left[  \frac{G^{(1,0,0)}_{s,s,s}}{2}  \left( \frac{1}{\da_c^2} + \frac{1}{S_s} \right) - \frac{\pd G^{(1,0,0)}_{s,s,s} }{ \pd S_s} \frac{S_s}{\da_c^2} \right]~\delta_l.
\label{EQ:SI_dh_2_dl}
\end{equation}
We note that $G^{(1,0,0)}_{s,s,s}$ can be rewritten as 

\begin{equation}
G^{(1,0,0)}_{s,s,s} = \frac{1}{3} \frac{\pd \langle \da_s^3 \rangle_c}{\pd S_s} = \frac{\mathcal{S}^{(3)}_s }{6} S_s \left( 2 \frac{\pd \ln \mathcal{S}_s^{(3)}}{ \pd \ln S_s } + 4 \right),
\end{equation}
and its derivative,

\begin{equation}
\frac{\pd G^{(1,0,0)}_{s,s,s}}{ \pd S_s} = \frac{\mathcal{S}^{(3)}_s}{6} \left(  8 \frac{\pd \ln \mathcal{S}^{(3)}_s}{ \pd \ln S_s} + 4+ \frac{2 S_s^2}{\mathcal{S}^{(3)}_s} \frac{\pd^2 \mathcal{S}^{(3)}_s}{\pd S_s^2} \right).
\end{equation}
Combining these equations with (\ref{EQ:SI_dh_2_dl}) gives equation (\ref{EQ:SI_bias_2}).

\subsection{Scale-dependent terms}

\subsubsection{Leading order}
\label{APP:SD_bias_1}

There are four terms in equation (\ref{EQ:dh_d1}) to consider:  

\begin{equation}
d_1 \equiv  \frac{\pd \langle \da_s^2 \da_l \rangle_c}{\da S_s}   \frac{\pd \fcollgm}{\pd S_s}\left(\frac{\pd \avgfcollgm}{\pd S_s} \right)^{-1} \psi 
\label{EQ:d1}
\end{equation}

\begin{equation}
 d_2 \equiv  \langle \da_s^2 \da_l \rangle_c ~ \frac{\pd \fcollgm}{\pd S_s}\left(\frac{\pd \avgfcollgm}{\pd S_s} \right)^{-1} \beta \psi
 \label{EQ:d2}
 \end{equation}
 
 \begin{equation}
 d_3 \equiv - 3 \frac{\pd \langle \da_s^2 \da_l \rangle_c}{\pd S_s}~ \frac{\pd \fcollgm}{\pd S_s}\left(\frac{\pd \avgfcollgm}{\pd S_s} \right)^{-1} \chi
 \label{EQ:d3}
\end{equation}

\begin{equation}
d_4 \equiv -  3 \langle \da_s^2 \da_l \rangle_c~ \frac{\pd \fcollgm}{\pd S_s}\left(\frac{\pd \avgfcollgm}{\pd S_s} \right)^{-1} \alpha \chi. 
\label{EQ:d4}
\end{equation}
We begin with equation (\ref{EQ:d2}).  Expanding $d_2$ to first order in $\da_l$ produces a constant, which will vanish when taking the cross correlation between $\da_h$ and $\da_l$, and a term proportional to $\da_l$.  Taking the cross-correlation of $\da_l$ with the term proportional to $\da_l$ gives

\begin{equation}
\langle d_2 \da_l \rangle = \langle \da_s^2 \da_l \rangle_c \left( \frac{\pd \avgfcollgm}{\pd S_s} \right)^{-1}  \lim_{S_l \rightarrow 0}  \left[ \frac{\pd }{\pd \da_l} \left. \left( \frac{ \pd\fcollgm}{\pd S_s} \beta \psi \right)\right|_{\da_l=0} \right] \left\langle \da_l\cdot \da_l \right\rangle =  \langle \da_s^2 \da_l \rangle_c \lim_{S_l\rightarrow 0}\left( \left. \beta  \right|_{\da_l=0}\right) \cdot  \lim_{S_l \rightarrow 0} \left[     \left. \frac{\pd \psi}{\pd \delta_l} \right|_{\da_l=0} ~S_l \right],
\label{EQ:limterm1}
\end{equation}
where we have used the fact that

\begin{equation}
 \lim_{S_l \rightarrow 0} \left. \psi \right|_{\da_l=0}= 0
 \label{EQ:convenient0}
\end{equation}
to obtain the right-hand side of the equation.  Conveniently, 

\begin{equation}
 \lim_{S_l \rightarrow 0} \left[  \left. \frac{\pd \psi}{\pd \delta_l} \right|_{\da_l=0} ~S_l \right] =1,
\label{EQ:convenient}
\end{equation}
so we are left with

\begin{equation}
\langle d_2 \da_l \rangle = \langle \da_s^2 \da_l \rangle_c \lim_{S_l\rightarrow 0}\left( \left. \beta \right|_{\da_l=0}\right) = \frac{\langle \da_s^2 \da_l \rangle_c}{2 S_s} \left( \frac{\da_c^2}{S_s} - 3\right).
\end{equation}
Now let us consider the contribution from (\ref{EQ:d1}).  Again, we expand about $\da_l=0$ to first order in $\da_l$ and take the cross-correlation with $\da_l$.  Making use of equations (\ref{EQ:convenient0}) and (\ref{EQ:convenient}), we obtain

\begin{equation}
\langle d_1 \da_l \rangle = \frac{\pd \langle \da_s^2 \da_l \rangle_c}{\pd S_s}  \lim_{S_l \rightarrow 0} \left( \left. \frac{\pd \psi}{\pd \delta_l} \right|_{\da_l=0} S_l \right)  = \frac{\pd \langle \da_s^2 \da_l \rangle_c}{\pd S_s},
\label{EQ:sdA}
\end{equation}
which can be rewritten as

\begin{equation}
\langle d_1 \da_l \rangle = \frac{\langle \da_s^2 \da_l \rangle_c}{S_s} + S_s \frac{\pd }{ \pd S_s}\left( \frac{\langle \da_s^2 \da_l\rangle_c}{S_s} \right).
\label{EQ:sdA2}
\end{equation}
It can be shown using the above approach that there are no contributions from $d_3$ and $d_4$.  Adding the contributions from $d_1$ and $d_2$ gives equation (\ref{EQ:sum_d1d2}).

\subsubsection{Next-to-leading order}
\label{APP:SD_bias_2}

From equation (\ref{EQ:dh_d2}), define: 
 
\begin{equation}
d_5 = 2 \frac{\pd G^{(0,1,0)}_{l,s,s}}{\pd S_s}  \left( \frac{\pd \avgfcollgm}{\pd S_s} \right)^{-1}  \frac{\pd \fcollgm}{\pd S_s} \mu
\end{equation}

\begin{equation}
d_6 = 3 G^{(0,1,0)}_{l,s,s} \left( \frac{\pd \avgfcollgm}{\pd S_s} \right)^{-1}  \frac{\pd \fcollgm}{\pd S_s} \chi
\end{equation}

\begin{equation}
 d_7 =  - \frac{\pd G^{(0,1,0)}_{l,s,s}}{\pd S_s} \left( \frac{\pd \avgfcollgm}{\pd S_s} \right)^{-1} \fcollgm \frac{\da_l}{S_l} 
 \end{equation} 
  
\begin{equation}
 d_8 = - G^{(0,1,0)}_{l,s,s} \left( \frac{\pd \avgfcollgm}{\pd S_s} \right)^{-1}  \frac{\pd \fcollgm}{\pd S_s}  \frac{\da_l }{S_l}.
\end{equation}
It can be shown that $d_5$ and $d_6$ do not contribute.  Consider the contribution from $d_7$.  It is straightforward to see that

\begin{equation}
\langle d_7 \delta_l \rangle =  - \left( \frac{\pd \avgfcollgm}{\pd S_s} \right)^{-1} \frac{\pd G^{(0,1,0)}_{l,s,s}}{\pd S_s} \lim_{S_l \rightarrow 0} \left[ \left. \frac{\pd \left(  \da_l \fcollgm \right) }{\pd \da_l}  \right|_{\da_l=0}  \right]  = - \left( \frac{\pd \avgfcollgm}{\pd S_s} \right)^{-1} \frac{\pd G^{(0,1,0)}_{l,s,s}}{\pd S_s} \avgfcollgm.
\label{EQ:d7}
\end{equation}
Noting that $G^{(0,1,0)}_{l,s,s} = (1/2)~\pd \langle \da_s^2 \da_l \rangle_c / \pd S_s$, we can re-write equation (\ref{EQ:d7}) as
\begin{equation}
\langle d_7 \delta_l \rangle = - \left( \frac{\avgfcollgm}{\pd S_s} \right)^{-1} \avgfcollgm \left[ \frac{\pd }{\pd S_s} \left( \frac{\langle \da_s^2 \da_l \rangle_c}{S_s} \right) + \frac{S_s}{2} \frac{\pd^2}{\pd S_s^2 } \left( \frac{\langle \da_s^2 \da_l \rangle)_c}{S_s} \right) \right].
\end{equation}
Expanding $d_8$ to first order in $\da_l$, taking the cross-correlation with $\da_l$, and the limit $S_s\rightarrow 0$, gives

\begin{equation}
\langle d_8 \delta_l \rangle =   -\left( \frac{\pd \avgfcollgm}{\pd S_s} \right)^{-1}  G^{(0,1,0)}_{l,s,s} \lim_{S_l \rightarrow 0} \left[ \left. \frac{\pd }{\pd \da_l}  \left(  \da_l \frac{\pd \fcollgm}{\pd S_s} \right)  \right|_{\da_l=0}  \right]  =  -G^{(0,1,0)}_{l,s,s},
\end{equation}
which can be written as

\begin{equation}
\langle d_8 \delta_l \rangle = -\frac{\langle \da_s^2 \da_l \rangle_c}{2 S_s} - \frac{S_s}{2}\frac{\pd }{\pd S_s}\left( \frac{\langle \da_s^2 \da_l \rangle_c}{S_s} \right).
\end{equation}
Adding the contributions from $d_7$ and $d_8$, and substituting the three-point correlator from (\ref{EQ:mixedcorrelator}) yields equation (\ref{EQ:sd_bias_2}).

\end{document}